**Title:** Integrated step selection analysis: bridging the gap between resource selection and animal movement

**Authors:**
1. Tal Avgar, Department of Biological Sciences, University of Alberta, Edmonton, Alberta, Canada, T6G 2G1.
2. Jonathan R. Potts, School of Mathematics and Statistics, University of Sheffield, Sheffield, UK.
3. Mark A. Lewis, Department of Biological Sciences and Department of Mathematical and Statistical Sciences, University of Alberta, Edmonton, Alberta, Canada, T6G 2G1
4. Mark S. Boyce, Department of Biological Sciences, University of Alberta, Edmonton, Alberta, Canada, T6G 2G1

**Corresponding author:** Tal Avgar, email: avgar@ualberta.ca

**Abstract:**
1. A resource selection function is a model of the likelihood that an available spatial unit will be used by an animal, given its resource value. But how do we appropriately define availability? Step-selection analysis deals with this problem at the scale of the observed positional data, by matching each 'used step' (connecting two consecutive observed positions of the animal) with a set of 'available steps' randomly sampled from a distribution of observed steps or their characteristics.
2. Here we present a simple extension to this approach, termed integrated step-selection analysis (iSSA), which relaxes the implicit assumption that observed movement attributes (i.e. velocities and their temporal autocorrelations) are independent of resource selection. Instead, iSSA relies on simultaneously estimating movement and resource-selection parameters, thus allowing simple likelihood-based inference of resource selection within a mechanistic movement model.
3. We provide theoretical underpinning of iSSA, as well as practical guidelines to its implementation. Using computer simulations, we evaluate the inferential and predictive capacity of iSSA compared to currently used methods
4. Our work demonstrates the utility of iSSA as a general, flexible and user-friendly approach for both evaluating a variety of ecological hypotheses, and predicting future ecological patterns.

**Introduction**

Ecology is the scientific study of processes that determine the distribution and abundance of organisms in space and time (Elton 1927). Hence, asking how and why living beings change their spatial position through time is fundamental to ecological research (Nathan *et al.* 2008). Animal movement links the behavioural ecology of individuals with population and community level processes (Lima & Zollner 1996). Its study is consequently vital for understanding basic ecological processes, as well as for applications in wildlife management and conservation.

Whether basic or applied, common to many empirical studies of animal movement is the aspiration to reliably predict population density through space and time by modelling the spatiotemporal probability of animal occurrence, also known as the utilization distribution (Keating & Cherry 2009). Despite much progress in statistical characterization of animal movement and habitat associations, our ability to predict utilization distributions is limited by our understanding of the underlying behavioural processes. Indeed, including explicit movement behaviours into spatial models of animal density has led to improved predictive performance (Moorcroft *et al.* 2006; Fordham *et al.* 2014). Deriving predictive space-use models based on the behavioural process underlying animal movement patterns is of particular importance when dealing with altered or novel landscapes that might differ substantially from the landscape used to inform the models.

Over the past three decades, a great deal of research has been dedicated to explaining and predicting spatial population distribution patterns based on underlying habitat attributes (often termed resources). In that regard, much focus has been given to estimating resource selection functions (Manly *et al.* 2002) - phenomenological models of the relative probability that an available discrete spatial unit (e.g., an encountered patch or landscape-pixel) will be used given its resource type/value (Lele *et al.* 2013). Indeed, its intuitive nature and ease of application has made resource selection analysis (RSA) the tool of choice for many wildlife scientists and managers seeking to use environmental information in conjunction with animal positional data (Boyce & McDonald 1999; McDonald *et al.* 2013; Boyce *et al.* 2015).

Whereas much progress has been gained in the application of RSAs to animal positional data, the problem of defining the appropriate spatial domain available to the animal remains as a major concern (Matthiopoulos 2003; Northrup *et al.* 2013; Lele *et al.* 2013; McDonald *et al.* 2013). Weighted distribution approaches deal with this problem by modelling space-use as a function of a movement model and a selection function, but most weighted distribution models are challenging to implement (but see Johnson *et al.* 2013). Matched case-control logistic regressions (CLRs; also known as discrete-choice models) may be considered a simplified alternative to the weighted distribution approach where each observed location is matched with a conditional availability set, limited to some predefined spatial and/or temporal range (Arthur *et al.* 1996; McCracken *et al.* 1998; Compton *et al.* 2002; Boyce *et al.* 2003; Baasch *et al.* 2010). A major strength of this approach is that maximum-likelihood estimates (MLEs) of the parameters can be efficiently obtained though commonly used statistical software (often relying on a Cox Proportional Hazard routine; e.g., function *clogit* in R). One particular type of such conditional RSA is step-selection analysis (SSA), where each 'used step' (connecting two consecutive observed positions of the animal) is coupled with a set of 'available steps' randomly sampled from the empirical distribution of observed steps or their characteristics (e.g., their length and direction; Fortin *et al.* 2005; Duchesne *et al.* 2010; Thurfjell *et al.* 2014).

The definition of availability is challenging however, even when using the SSA approach. The problem arises due to the sequential, rather than simultaneous, estimation of the movement and habitat-selection components of the process. Owing to this stepwise procedure, the resulting habitat-selection inference is conditional (on movement), whereas movement is assumed independent of habitat selection. In reality, the two are tightly linked, with habitat selection and availability affecting the animal's movement patterns (Avgar *et al.* 2013b), and the animal's movement capacity affecting its habitat-use patterns (Rhodes *et al.* 2005; Avgar *et al.* 2015). Failure to adequately account for the movement process may consequently lead to biased habitat selection estimates (Forester *et al.* 2009).

As we will show here, the benefits of adequately accounting for the movement process may extend beyond obtaining unbiased habitat selection estimates. SSAs rely on a simple depiction of animal movement as a series of stochastic discrete steps, characterised by specific velocity and autocorrelation distributions. This same depiction underlies the mathematical modelling of animal movement as a discrete-time random walk (RW), including correlated and/or biased RW (Kareiva & Shigesada 1983; Turchin 1998; Codling *et al.* 2008). Indeed, many SSA formulations correspond to a correlated RW process with local bias produced by resource selection (BCRW; Duchesne *et al.* 2015). Apart from their compatibility with the way we often observe animal movement (i.e. in continuous space and at discrete times), many RW can be well approximated by diffusion equations, allowing a much sought shift from an individual-based Lagrangian perspective to population-level Eulerian models (Turchin 1991, 1998). SSAs are thus at an interface between statistical (phenomenological) RSAs and mathematical (mechanistic) RW models (Potts *et al.* 2014b), models that form the backbone of much of the existing body of theory in the field of animal movement (Codling *et al.* 2008; Fagan & Calabrese 2014; Benhamou 2014).

In this paper we outline a CLR-based approach for simultaneous estimation of the movement and habitat-selection components, an approach we name *integrated step-selection analysis* (iSSA; Fig 1). The iSSA allows the effects of environmental variables on the movement and selection processes to be distinguished, thus providing a valuable tool for testing hypotheses (e.g., to test whether animals travel faster in certain times or through certain habitats), while resulting in an empirically parameterised mechanistic movement model (i.e., a mechanistic step-selection model; Potts *et al.* 2014a), that can be used to translate individual-level observations to population-level utilisation distributions across space and time (Potts *et al.* 2014b; a; Appendix S1).

The iSSA is related to several recently published works integrating animal movement and resource selection. Both Forester *et al.* (2009) and Warton & Aarts (2013) demonstrated the inclusion of movement variables in an RSA and its marked effect on the resulting inference. Johnson *et al.* (2013) have shown that animal telemetry data can be viewed as a realisation of a non-homogenous space–time point process, and MLEs of this process can be obtained using a generalised linear model. These contributions focused on gaining unbiased resource selection inference while treating the movement process as nuisance that must be 'controlled for'. Here, we seek explicit inference of this process. State-space models of animal movement (reviewed by Jonsen *et al.* 2003 and Patterson *et al.* 2008) predict the future state (e.g., spatial position) of the animal given its current state (where an 'observation model' provides the probability of observing these states), environmental covariates, and an explicit 'process model'. Once parametrised, the process model can be used to generate space-use prediction, but parametrisation is often technically demanding and computationally intensive (Patterson *et al.* 2008). More recently, Potts *et al.* (2014a) demonstrated the use of a 'mechanistic step-selection model' to infer both the drivers and the steady-state distribution of animal space-use, but the model was framed around one specific functional form of the movement kernel, and parameter estimates were obtained using a custom-made likelihood maximisation procedure. Lastly, Duchesne *et al.* (2015) demonstrated that an SSA can be used to obtain unbiased estimates of the directional persistence and bias of a BCRW, but did not address parametrisation of the step-length distribution.

The iSSA builds and expands on these contributions. We will demonstrate that, by statistically accounting for an explicit movement process within an SSA, a complete habitat-dependent mechanistic movement model can be parametrised from telemetry data using a standard CLR routine. In the following we provide a detailed description of the approach, and evaluate its performance (compared with standard RSA and SSA) in correctly inferring the movement and habitat-selection processes underlying observed space-use patterns, and in predicting the resulting UD.

**Materials and methods**

Integrated Step-Selection Analysis: in their work on the subject of accounting for movement in resource-selection analysis, Forester *et al.* (2009) demonstrated that including a distance function in SSA substantially reduces the bias in habitat-selection estimates. Mathematically, their argument is based on the habitat-independent movement kernel (the function governing movement in the absence of resource selection, or across a homogeneous landscape; Hjermann 2000; Rhodes *et al.* 2005) belonging to the exponential family, so that it can be accounted for with the logistic formulation of the SSA likelihood function. Here we shall make the assumption that, in the absence of habitat selection, step lengths follow either an exponential, half-normal, gamma, or log-normal distribution. Under this assumption, the statistical coefficients associated with step length, its square, its natural logarithm, and/or the square of its natural logarithm (depending on the assumed distribution), when incorporated as covariates in a standard SSA, serve as statistical estimators of the parameters of the assumed step-length distribution (see Appendices S2-S3 for details, and below for an example). Standard model selection (e.g., likelihood-ratio or AIC) then can be used to select the best-fit theoretical distribution (out of the four listed above).

The iSSA approach moreover can be applied to infer directional persistence and external bias. Assuming that the angular deviations from preferred directions (either the previous heading, the target heading, or both) are von Mises distributed (an analog of the normal distribution on the circle), the cosine of these angular deviations can be included as covariates in an SSA to obtain MLEs of the corresponding von Mises concentration parameters (Duchesne *et al.* 2015). Hence, MLEs of iSSA coefficients affiliated with directional deviations and step-lengths are directly interpretable as the parameters of distributions governing the underlying BCRW.

We shall make the assumption here that animal space-use behaviour is adequately captured by a separable model, involving the product of two kernels, a movement kernel and a habitat-selection kernel. Formally, we define a discrete-time movement kernel, $\Phi$, which is proportional to the probability density of occurrence in any spatial position, $x$, at time $t$, in the absence of habitat selection. The determinants of $\Phi$ are: the Euclidian distances between $x$ and the preceding position, $x_{t-1}$ (the step length; $l_t = ||x - x_{t-1}||$), the distances between $x_{t-1}$ and $x_{t-2}$ (the previous step length; $l_{t-1} = ||x_{t-1} - x_{t-2}||$), the associated step headings, $\alpha_t$ and $\alpha_{t-1}$ (the directions of movement from $x_{t-1}$ to $x$ and from $x_{t-2}$ to $x_{t-1}$, respectively), and a vector of spatial and/or temporal movement predictors at time $t$ and/or at the vicinity of $x$ and/or $x_{t-1}$, $Y(x,x_{t-1},t)$ (e.g., terrain ruggedness, migratory phase, snow depth, etc.). The effects of these step attributes on $\Phi$ are controlled by the associated coefficients vector, $\theta$. Note that the effects of spatial attributes here are assumed to operate through local biomechanical interactions between the animal and its immediate environment, interactions that determine the rate of displacement (i.e., kinesis), not where the animal 'wants' to be (i.e., taxis). Also note that the kernel $\Phi$ can be non-Markovian and accommodate various types of velocity autocorrelations (lack of independence between directions and/or lengths of consecutive steps), including correlated and biased random walks (if directional biases are known a priori).

We further define the habitat-selection function, $\Psi$, which is proportional to the probability density of observing the animal in any spatial position, $x$, at time $t$, in the absence of movement constraints. The determinants of $\Psi$ are the habitat attributes in $x$ at $t$, $H(x,t)$, and their corresponding selection coefficients, $\omega$. The normalised product of $\Phi$ and $\Psi$ yields the probability density of occurrence in $x$ at $t$, which is:

$$f(x_t|x_{t-1}, x_{t-2}) = \frac{\Phi[l_t, l_{t-1}, \alpha_t, \alpha_{t-1}, Y(x, x_{t-1}, t), \theta] \cdot \Psi[H(x,t), \omega]}{\int_\Omega \Phi[l_t, l_{t-1}, \alpha_t, \alpha_{t-1}, Y(x, x_{t-1}, t), \theta] \cdot \Psi[H(x,t), \omega] dx} . \qquad \text{Eq. 1}$$

Note that the same environmental variable (e.g., snow-depth or terrain ruggedness) might be included in both $Y$ and $H$ and hence affect both $\Phi$ (e.g., decreased speed in deep snow or across rugged terrain) and $\Psi$ (e.g., selection for snow-free or flat localities). Eq. 1 is equivalent to the formulation used (for example) by Rhodes *et al.* (2005, Eq. 1), Forester *et al.* (2009, Eq. 1), and Johnson *et al.* (2013, Eq. 1), and is a generalised form of a redistribution kernel – a widely used mechanistic model of animal movement and habitat selection (see Discussion for recent examples).

The denominator in Eq. 1 is an integral over the entire spatial domain, $\Omega$, serving as a normalisation factor to ensure the resulting probability density integrates to one. Whereas in most cases it would be impossible to solve this integral analytically, various forms of numerical (discrete-space) approximations can be used to fit redistribution kernel functions, such as Eq. 1, to data (see Avgar *et al.* 2013a and the Discussion). Here we focus on a simple likelihood-based alternative to such numerical methods, one that can be implemented using common statistical software and is hence accessible to most ecologists. Assuming an exponential form for both $\Phi$ and $\Psi$, MLEs for the parameter vectors $\theta$ and $\omega$ can be obtained using conditional logistic regression, where observed positions (cases) are matched with a sample of available positions (controls; Fig 1 and Appendices S2-S4).

<u>A hypothetical example:</u> Let us assume we have obtained a set of $T$ spatial positions, sampled at a unit temporal interval along an animal's path, and that we also have maps of two (temporally stationary) spatial covariates, $h(x)$ and $y(x)$. We shall now assess the statistical support for the following propositions (examples of the sort of hypotheses that could be tested):

A. The animal is selecting high values of $h(x)$.
B. At the observed temporal scale, and in the absence of variability in $h(x)$, the animal's movement is directionally persistent (i.e., consecutive headings are positively correlated), and the degree of this persistence varies with $y(x)$ (e.g., the animal moves more directionally where $y(x)$ is lower). The resulting turn-angles are von Mises distributed with mean 0 (i.e., left and right turns are equally likely) and a $y$-dependent concentration parameter.
C. At the observed temporal scale, and in the absence of variability in $h(x)$ and $y(x)$, the animal's movement is characterised by gamma distributed step lengths, and the shape of this step-length distribution depends on the time of day (e.g., the animal moves faster during daytime).

Note that these propositions are contingent on the temporal gap between observed relocations (i.e. step duration), as well as on the spatial resolution of our covariate maps, $h(x)$ and $y(x)$. We thus explicitly acknowledge that our inference is scale dependent.

We start by sampling, for each (but the first two) of the observed points along a path ($x_t$, $t$ = 3, 4, ..., $T$), a set of $s$ control points (available spatial positions at time $t$; $x'_{t,i}$, $i$ = 1, 2, ..., $s$), where the probability of obtaining a sample at some distance, $l'_{t,i}$, from the previous observed point ($l'_{t,i} = \|x'_{t,i} - x_{t-1}\|$) is given by the gamma PDF:

$$g(l'_{t,i}|b_1, b_2) = \frac{1}{\Gamma(b_1) \cdot b_2^{b_1}} \cdot {l'_{t,i}}^{b_1-1} \cdot e^{-\frac{l'_{t,i}}{b_2}}. \qquad \text{Eq. 2}$$

Here, $b_1$ and $b_2$ are initial estimates of the gamma shape and scale parameters (respectively) obtained based on the observed step-length distribution (using either the method of moments or maximum likelihood). As noted earlier, this estimation is confounded by the process of habitat selection and hence a method to unravel movement inference from habitat selection is needed. The iSSA will provide estimates of the deviations of these initial values from the unobserved habitat-independent shape and scale (Appendices S2-S3). Note that these control sets also could be sampled randomly within some finite spatial domain (e.g., within the maximal observed displacement distance; Appendices S2 and S4). Distance weighted sampling is expected to increase inferential efficiency, resulting in a smaller standard error for a given $s$ value, but is not a necessity (Forester *et al.* 2009). ). In general, any increase in $T$ and/or $s$ will result in better approximation of the used- and/or availability-distributions (respectively), and hence better inference (together with larger computational costs).

Once sampled, control (available) points, $x'_{t,i}$, are assigned a value of 0 whereas the observed (used or case) points, $x_t$, are assigned a value of 1. The resulting binomial response variable can now be statistically modeled using conditional (case-control) logistic regression, as the likelihood of the observed data is exactly proportional to (Gail *et al.* 1981; Forester *et al.* 2009; Duchesne *et al.* 2015):

$$\prod_{t=3}^{T} \frac{\exp[b_3 \cdot h(x_t) + [b_4 + b_5 \cdot y(x_{t-1})] \cdot \cos(\alpha_{t-1} - \alpha_t) + b_6 \cdot l_t + (b_7 + b_8 \cdot D_t) \cdot \ln(l_t)]}{\sum_{i=0}^{S} \exp[b_3 \cdot h(x'_{t,i}) + [b_4 + b_5 \cdot y(x_{t-1})] \cdot \cos(\alpha_{t-1} - \alpha'_{t,i}) + b_6 \cdot l'_{t,i} + (b_7 + b_8 \cdot D_t) \cdot \ln(l'_{t,i})]}, \quad \text{Eq. 3}$$

where $\alpha'_{t,i}$ is the direction of movement from $x_{t-1}$ to $x'_{t,i}$, and $D_t$ is an indicator variable having the value 1 when *t* is daytime and 0 otherwise. Note that the summation in the denominator starts at *s* = 0 (rather than 1) to indicate that the used step is included in the availability set ($x'_{t,i=0} = x_t$). Also note that it is the inclusion of turn-angles that necessitates the exclusion of the first two positions ($x_{t=1}$ and $x_{t=2}$); if no velocity autocorrelation is modeled, only the first position is excluded. Lastly, note that this formulation implies that the degree of directional persistence is affected by the value of *y* at the onset of the step only; in the next section we provide an example of modelling habitat effects on movement along the step.

Eq. 3 is a discrete-choice approximation of Eq. 1 (specifically tailored according to propositions A-C) and we provide its full derivation in Appendix S3. In summary, $b_3$ is the habitat selection coefficient (corresponding to proposition A and estimating the only component of the parameter vector *ω* in Eq. 1), $b_4$ and $b_5$ are the basal (habitat-independent) and *y*-dependent directional persistence coefficients (corresponding to proposition B and estimating two components of the parameter vector *θ* in Eq. 1), and $b_6$, $b_7$, and $b_8$ are the modifiers of the step-length shape and scale coefficients (corresponding to proposition C and estimating the remaining components of the parameter vector *θ*). Once maximum-likelihood estimates are obtained, the shape and scale parameters of the basal step-length distribution can be calculated (Appendix S3), where the shape is given by: $[(b_1 + b_7) + b_8 \cdot D_t]$, and the scale is given by: $[1/(b_2^{-1} - b_6)]$. Similarly, $b_4$ can be shown to be an unbiased estimator of the concentration parameter of the (habitat-independent) von Mises turn-angles distribution (Duchesne *et al.* 2015).

Including movement attributes as covariates in SSA, which we termed here iSSA, thus allows simple likelihood-based estimation of explicit ecological hypotheses within a framework of a mechanistic habitat-mediated movement model. Such hypotheses might include, in addition to those mentioned thus far, long- and short-term target prioritisation (Duchesne *et al.* 2015), barrier crossing and avoidance behaviour (Beyer *et al.* 2015), and interactions with conspecifics and intraspecifics (Latombe *et al.* 2014; Potts *et al.* 2014b). In fact, many of the facets of the generic approach developed by Langrock *et al.* (2012) can be included in an iSSA with the MLEs obtained using standard statistical packages. An iSSA thus holds promise as a user-friendly yet versatile approach in the movement ecologist's toolbox. In Appendix S4 we provide practical guidelines for the application of iSSA. In the next sections we explore the utility of this approach using computer simulations.

Simulations: Testing the inferential and predictive capacities of any statistical space-use model is challenging because we are often ignorant of the true process giving rise to the observed patterns, as well as of the true distribution of space-use from which these patterns are sampled (Avgar *et al.* 2013a; Van Moorter *et al.* 2013). To deal with this challenge we employ here a simple process-based movement simulation framework. We provide full details of the simulation procedure and its statistical analysis in Appendix S5.

Fine-scale space-use dynamics were simulated using stochastic 'stepping-stone' movement across a hexagonal grid of cells. Each cell, $x$, is characterised by habitat quality, $h(x)$ with spatial autocorrelation set by an autocorrelation range parameter, $\rho$ (= 0, 1, 5, 10, and 50). For each $\rho$ value, 1,000 trajectories were simulated and then rarefied (by retaining every $100^{th}$ position). Each of these rarefied trajectories was then separately analysed using RSA and 10 different (i)SSA formulations, including one or more of the following covariates (Table 1): habitat values at the end of each step, $h(x_t)$, the average habitat value along each step, $h(x_{t-1},x_t)$, the step-length, $l_t$ (= $||x_{t-1} - x_t||$), its natural-log transformation, $ln(l_t)$, and the statistical interactions between $l_t$, $ln(l_t)$, and $h(x_{t-1},x_t)$. Models that included only $h(x_t)$ and/or $h(x_{t-1},x_t)$ correspond to traditionally used SSA (relying on empirical movement distributions with no movement attribute included as covariates; models *a*, *b*, and *c* in Table 1), whereas models that additionally included $l_t$ and $ln(l_t)$ correspond to iSSA. The predictive capacity of the models was estimated based on the agreement between their predicted utilisation distributions (UD) and the 'true' UD, generated by the true underlying movement process. We refer the reader to Appendix S5 for further details.

A separate simulation study was conducted to evaluate the identifiability and estimability of the iSSA parameters as function of sample size and habitat selection strength (Appendix S6).

**Results**

Parameterization: All models converged in a timely manner and the convergence time for the most complex model (model *j* in Table 1) was approximately 1 CPU sec. Of the 10 (i)SSA formulations specified in Table 1, AIC ranking indicated support for only four (*d*, *f*, *h*, and *j*), all of which include the habitat value at the step's endpoint (with coefficient $\beta_3$) and the step-length and its natural logarithm (with coefficients $\beta_5$ and $\beta_6$) as covariates. Hence, iSSAs better explain our simulated data than traditionally used SSAs (excluding step length as a covariate), but only as long as an endpoint effect (i.e., selection for/against the habitat value at the end of the step) is included. In fact, models that excluded the habitat value at the step's endpoint (models *b*, *e*, *g*, and *i*) had AIC scores that were typically two orders of magnitude larger than those including it. In comparison to RSA, iSSA formulations had unequivocal AIC support at low habitat spatial autocorrelation levels, but only partial support at high autocorrelation levels (Table 1).

Estimated habitat selection strengths, as indicated by our RSA and SSA coefficient estimates ($\beta_{RSA}$ and $\beta_3$ respectively), were appreciably larger than the true habitat selection strength ($\omega = 1$), and more so for RSA estimates than for SSA (Fig. 2). Note that this in itself does not mean these estimates are 'biased' but rather reflects the inherent difference between the intensity of the true process and that of the emerging pattern, at the scale of observation (see further discussion below). These estimates showed little sensitivity to the level of habitat spatial autocorrelation, although a substantial increase in variance is observed in the RSA case (Fig. 2a). As found before by Forester *et al.* (2009), the strength of SSA-inferred habitat selection is larger when step lengths are included as a covariate in the analysis (iSSA), but this effect is fairly weak and diminishes as the habitat's spatial autocorrelation increases (Fig. 2b).

Overall, SSA-inferred habitat selections were substantially less variable than RSA-based estimates and showed little sensitivity to the inclusion or exclusion of other covariates in the model fit (Fig 2). This is not the case, however, for the effect of the mean habitat value along the step ($\beta_4$), which varied substantially with both the level of habitat spatial autocorrelation and the inclusion of an endpoint effect ($\beta_3$). Where $\beta_3$ was not included in the model fit (models *b*, *e*, and *i* in Table 1), $\beta_4$ increased with $\rho$, whereas where $\beta_3$ was included (models *c*, *f*, and *j*), $\beta_4$ was closer to zero (Fig. 3). Interestingly, when only the habitat at the end of the step and the habitat along the step were included in the model (i.e. model *c*; a commonly used SSA formulation), and at low $\rho$ values (= 0, 1), $\beta_4$ was negative, indicating 'selection against' high-quality steps. In fact, this reflects the low probability of observing a 'used' step that traverses high-quality habitat but does not end there.

As explained above (and in Appendices S2-S3), iSSA coefficients affiliated with the step length ($\beta_5$) and its natural logarithm ($\beta_6$), when combined with the estimated shape and scale values of the observed step-length distribution ($\beta_1$ and $\beta_2$; on which sampling was conditioned; Appendices S3 and S5), could be used to infer the shape and scale of the 'habitat-independent' step-length distribution [i.e., assuming $h(x_t, x_{t+1}) = 0$]. Under most imaginable scenarios, we would expect this basal movement kernel to be wider (i.e., with larger mean) than the observed one, as animals tend to linger in preferred habitats and hence display more restricted movements compared to the basal expectation. Indeed, the mean of these inferred distributions (the product of their shape and scale: $\frac{\beta_1+\beta_6}{\beta_2^{-1}-\beta_5}$) corresponds exactly to the observed mean, as long as no other covariates are included in the analysis (model *x* in Fig. 4). Once other covariates are included in the model (and hence habitat selection is at least partially accounted for), inferred mean step-length values were significantly higher from the observed values, showed little sensitivity to model structure, but increased with $\rho$ (as do the observed mean step-lengths). One exception is model *g*, which strongly underestimated the mean-step length at moderate-high $\rho$ values as it does not include any main habitat effects.

Even at high $\rho$ values, inferred mean step-length slightly but consistently underestimates the true habitat-independent step-length distribution (calculated by simulating the process based on Eq. S5.1 with $\omega = 0$; Fig. 4). This bias is a result of an iSSA's limited capacity to account for the full movement process as it unfolds in between observations. The animal does not actually travel along the straight lines that we term 'steps' and, even if it would, the mean habitat value along the step does not exactly correspond to its probability to travel farther. As long as the scale of the observation is coarser than the scale of the underlying movement process, the animal's true movement capacity is never fully manifested in the observed relocation pattern and is thus always underestimated. Note however that this bias is negligibly small where the spatial autocorrelation of habitats is high ($\rho > 1$; Fig. 4).

Finally, despite apparent support for iSSA formulations including interaction between the step length and habitat quality along the step (Table 1; models *h* and *j*), the estimated values of the interaction coefficients, $\beta_7$ and $\beta_8$, mostly overlapped zero (Appendix S7). Generally speaking, the mean habitat value along the step has a weak negative effect on both the shape (through $\beta_8$) and the scale (through its inverse relationship with $\beta_7$) of the step-length distribution – long steps are less likely through high-quality habitats.

<u>Predictive capacities:</u> At approximate steady-state, RSA-based UD predictions are slightly more accurate and precise than SSA-based predictions (Fig 5 and Appendix Table S8). The RSA's predictive capacity increases with $\rho$ (while its precision dramatically decreases), whereas the opposite is true for SSA predictions, where the minimum *KLD* value (Kullback-Leibler Divergence; see Appendix S5) is reached when $\rho = 0$ (Appendix Table S8). *KLD* values coarsely mirror the AIC ranking of the different SSA formulations in distinguishing those that include an endpoint effect ($\beta_3$), but the best performing formulations based on *KLD* are simpler than the ones selected based on AIC (Tables 1 and S8). That said, all iSSA formulations including an endpoint effect performed well overall, with $G_{KLD}$ scores (a measure of goodness of fit; Appendix S5) ranging from ~0.84 (model *f* when $\rho = 50$) to ~0.98 (model *d* when $\rho = 0$). For reference, the $G_{KLD}$ scores for RSA-based predictions ranged from ~0.98 ($\rho = 0$) to ~0.99 ($\rho = 50$).

To test the sensitivity of the models' predictions to the sampling scale (see Appendix S9 for relating $\rho$ to the sampling scale), we generated predicted UDs, both SSA-based and RSA-based, across a highly autocorrelated landscape ($\rho = 50$) using parameter estimates obtained from samples of a random landscape ($\rho = 0$), and vice versa. RSA-based predictions were robust to these scale-mismatches, with $G_{KLD}$ scores of ~0.98 and ~0.96, for the $\rho = 50$ landscape (with parameter estimates based on $\rho = 0$ data) and the $\rho = 0$ landscape (with parameter estimates based on $\rho = 50$ data), respectively. Similarly, all step-selection models including an endpoint effect performed well, with $G_{KLD}$ scores ranging from ~0.94 (model *f*) to ~0.98 (model *h*) for the $\rho = 0$ landscape (with parameter estimates based on $\rho = 50$ data), and $G_{KLD}$ scores ranging from ~0.83 (model *j*) to ~0.97 (model *a*) for the $\rho = 50$ landscape (with parameter estimates based on $\rho = 0$ data). Overall, iSSA-based predictive capacity remained mostly unaltered by mismatches between the data's landscape structure and the structure of the landscape on which projections are made.

As can be expected, step-selection models predict transient UDs better than the inherently stationary RSA (except when $\rho = 50$; Appendix Table S8). In comparison with steady-state predictions, complex iSSA-based predictions perform better than simpler ones (Appendix Table S8). As for the steady-state predictions, all iSSA formulations including an endpoint effect performed well in predicting transient UDs, with $G_{KLD}$ scores ranging from ~89% (model *h* when $\rho = 0$) to ~0.98 (model *d* when $\rho = 1$). $G_{KLD}$ scores for RSA-based predictions showed substantial sensitivity to the level of spatial autocorrelation, ranging from ~0.69 ($\rho = 0$) to ~0.97 ($\rho = 50$).

<u>iSSA identifiability and estimability:</u> Results are presented in detail in Appendix S6. In short, our analysis revealed that, under the test-scenario, all iSSA parameters are fully identifiable, that estimates are unbiased in relation to the true values of the kernel generating functions, and that an increase in sample size beyond ~400 observed positions does not seem to substantially enhance precision (and hence estimability). That said, our results also indicate that inferential accuracy of movement related parameters may be highly variable, leading to compromised precision (with up to 1000% departure from the true value; Appendix S6) even at a fairly large sample size. This may be particularly true given the inherent trade-off between sampling extent and frequency (Fieberg 2007). Estimability of certain parameters, under certain scenarios, may thus be weak and must be evaluated on a case-by-case basis.

**Discussion**

The ideas, simulations and results presented above are aimed at providing a comprehensive assessment of using integrated step selection analysis, iSSA, with emphasis on its predictive capacity. The iSSA allows simultaneous inference of habitat-dependent movement and habitat-selection and is hence a powerful tool for both evaluating ecological hypotheses and predicting ecological patterns. We have shown that iSSA-based habitat selection inference is relatively insensitive to model structure and landscape configuration, and that iSSA-based UD predictions perform well across different temporal and spatial scales (we discuss the connection between the temporal resolution of the data and the habitat spatial autocorrelation in Appendix S9). On the other hand, our results indicate that movement and habitat selection may not be completely separable once observations are collected at a coarser temporal resolution than the underlying behavioural process. Consequently, stationary RSA-based predictions, whereas much simpler to obtain, provide slight but consistent better fit to the true UD when the time scale is long (and hence approaches the steady-state limit).

Two caveats are in place here. First, in our analysis the definition of the availability set for the RSA was exact (i.e., the entire domain), a situation that seldom occurs in empirical studies where availability is unknown. This is not the case for iSSA where the availability set always can be adequately defined (but is conditional on the temporal resolution of the positional data). Second, the high variability characterising the RSA coefficient estimates, and its resulting predictions (Figs 2 and 5) indicate substantial risk of erroneous inference. This may be particularly true when sample size is smaller than the relatively large sample used here, resulting in data that are not adequate unbiased samples of the steady-state UD, which is likely the case in most empirical studies. The more mechanistic nature of the iSSA makes it less sensitive to stochastic differences between specific realisations of the space-use process (e.g., due to differences in landscape configuration), and thus leads to more precise inference. Hence, even if the sole objective of a given study is to predict the long-term (steady-state) utilisation distribution, the more complicated iSSA-based predictions might be more reliable than those based on RSA. Moreover, in many real-world ecological scenarios a steady-state is never reached, and consequently the static RSA-based approach is less appropriate than the dynamical iSSA. We thus conclude that iSSA should be the method of choice whenever: (1) RSA availability cannot be properly defined, (2) predicting across a landscape different from the landscape used for parametrisation, (3) the data used for parametrisation are not an adequate sample of the true steady-state UD, or (4) predicting transient space-use dynamics.

Many movement and selection processes could be considered plausible, and the particular details of the mechanistic model used to simulate space-use data might substantially alter our conclusions. Our aim here was to use the simplest, and hence most general, mechanistic process imaginable, leading us to choose a stepping-stone movement process as our pattern-generating process. Interestingly, this simple process, governed by only two parameters (Eq. S5.1), gave rise to complex patterns once rarified. In particular, the emerging step-length distributions fit remarkably well with a gamma distribution, with shape and scale that reflect the underlying landscape structure. Note that this is a purely phenomenological description of the movement kernel, as the true underlying process had a fixed, habitat-independent movement parameter (Appendix S5). Ideally, a truly mechanistic approach will involve maximizing the likelihood over all possible paths the animal might have taken between two observed locations, and hence allowing inference of the true underlying process (Matthiopoulos 2003). In most cases however, this approach is forbiddingly computationally expensive. We showed that the approximation based on samples of straight-line movements between observed positions, which is the underlying assumption of any SSA, performs well over a range of conditions. An iSSA thus provides a reasonable compromise between computationally intensive mechanistic models and the purely phenomenological RSA.

According to Barnett & Moorcroft (2008), the steady-state UD should scale linearly with the underlying habitat-selection function $\Psi$ (Eq. 1) when informed movement capacity exceeds the scale of spatial variation in $\Psi$, but should scale with the square of $\Psi$ if informed movement capacity is much shorter than the scale of habitat variation. In the particular case of the exponential habitat selection function used here (Eq. S5.1), we would thus expect the following log-linear relationship: $ln[UD(x)] = a + b \cdot \omega \cdot h(x)$, where $a$ is a scaling parameter [the utilisation probability where $h(x) = 0$], and $b$ ($1 \leq b \leq 2$) is some increasing function of the habitat spatial autocorrelation, $\rho$. Our results, emerging from a very different model than the continuous-space continuous-time analytical approximation of Barnett & Moorcroft (2008), corroborate this expectation. The slope of the log-linear regression model described above increases from $b \approx 1.4$ to $b \approx 2$ as $\rho$ increases from 0 to 50 (Appendix S10). RSA-based coefficient estimates, $\beta_{RSA}$, closely mirror this pattern, increasing from ~1.6 to ~2 as $\rho$ increases (Fig 2a). Hence, as can be expected from a phenomenological model, RSA-based inference reflects the steady-state UD rather than the underlying habitat-selection process.

Recent years have seen a proliferation of sophisticated modelling approaches aimed at mechanistically capturing animal space-use behaviours. Many of these models share the theoretical underpinning of iSSA (as formulated in Eq. 1), relying on a depiction of animal space-use as emerging from the product of a resource-selection process and a selection-independent movement kernel (e.g., Rhodes *et al.* 2005; Getz & Saltz 2008; Avgar *et al.* 2013; Potts *et al.* 2014a; Beyer *et al.* 2015). Unlike the iSSA however, fitting these kernel-based models to empirical data relies on complex, and often specifically tailored likelihood-maximisation algorithms (namely discrete-space approximations of the integral in Eq. 1). The statistical machinery used in iSSA, based on obtaining a small set of random samples from an inclusive availability domain, is accessible to most ecologists because it relies on software that is already used (Thurfjell *et al.* 2014). Through the addition of appropriate covariates and interaction terms, iSSA can moreover address many of the questions that were the focus of other kernel-based approaches, such as home-range behaviour (Rhodes *et al.* 2005), memory-use (Avgar *et al.* 2013a, 2015; Merkle *et al.* 2014; Schlägel & Lewis 2015), habitat-dependent habitat selection (Potts *et al.* 2014a), and barrier effects (Beyer *et al.* 2015). Hence, iSSA allows ecologists to tackle complicated questions using simple tools.

To conclude, our work complements several recent contributions advocating the use of movement covariates within step-selection analysis (Forester *et al.* 2009; Johnson *et al.* 2013; Warton & Aarts 2013; Duchesne *et al.* 2015). We believe a convincing body of theoretical evidence now indicates the suitability of integrated step-selection analysis as a general, flexible and user-friendly approach for both evaluating ecological hypotheses and predicting future ecological patterns. Our work highlights the importance of including endpoint effects in the analysis together with some caveats regarding the interpretation of SSA results, specifically when dealing with the effects of the habitat along the step. We also recommend careful consideration of parameter estimability, particularly with regards to the movement components of the model, which may be prone to strong cross-correlations (as discussed in Appendix S6). Based on our current experience in applying iSSA to empirical data (Avgar, work in progress) we have provided practical guidelines in Appendix S4. Additional theoretical work is needed to investigate the effects of the underlying movement process on iSSA performance, as well as to come up with computationally efficient iSSA-based simulations to enable rapid generation of predicted utilisation distribution (as discussed in Appendix S1). Most importantly, the utility of iSSA must now be evaluated by applying it to real datasets, and using it to solve real ecological problems.

**Acknowledgments**
TA gratefully acknowledges supported by the Killam and Banting postdoctoral fellowships. MAL gratefully acknowledges NSERC Discovery and Accelerator Grants and a Canada Research Chair. MSB thanks NSERC and the Alberta Conservation Association for funding. The authors thank L. Broitman for designing Fig. 1 and C. Prokopenko for helpful editorial comments, and Dr. Geert Aarts, Dr. John Fieberg, and an anonymous reviewer, for their instructive comments and suggestions.

Table 1 – the 11 different models fitted here and their relative performance ranking at five different levels of habitat spatial autocorrelation (with 1000 realizations at each level). To enable AIC comparison, RSA's were run with only those positions included in the SSA (i.e., excluding the first position).

| model | | covariates | | | | | | % scored as best (based on AIC) | | | | |
|---|---|---|---|---|---|---|---|---|---|---|---|---|
| | | $R(x_t)$ | $R(x_t, x_{t-1})$ | $l_t$ | $ln(l_t)$ | $R(x_t, x_{t-1}) \cdot l_t$ | $R(x_t, x_{t-1}) \cdot ln(l_t)$ | $\rho = 0$ | $\rho = 1$ | $\rho = 5$ | $\rho = 10$ | $\rho = 50$ |
| RSA | | $\beta_{RSA}$ | 0 | 0 | 0 | 0 | 0 | 0 | 0 | 20.3 | **45.7** | 44.7 |
| SSA | a | $\beta_3$ | 0 | 0 | 0 | 0 | 0 | 0 | 0 | 0 | 0 | 0 |
| | b | 0 | $\beta_4$ | 0 | 0 | 0 | 0 | 0 | 0 | 0 | 0 | 0 |
| | c | $\beta_3$ | $\beta_4$ | 0 | 0 | 0 | 0 | 0 | 0 | 0 | 0 | 0 |
| iSSA | d | $\beta_3$ | 0 | $\beta_5$ | $\beta_6$ | 0 | 0 | 0 | 0 | 0 | 1.1 | 1.1 |
| | e | 0 | $\beta_4$ | $\beta_5$ | $\beta_6$ | 0 | 0 | 0 | 0 | 0 | 0 | 0 |
| | f | $\beta_3$ | $\beta_4$ | $\beta_5$ | $\beta_6$ | 0 | 0 | 10.4 | 4.4 | 13.9 | 24.9 | 28.4 |
| | g | 0 | 0 | $\beta_5$ | $\beta_6$ | $\beta_7$ | $\beta_8$ | 0 | 0 | 0 | 0 | 0 |
| | h | $\beta_3$ | 0 | $\beta_5$ | $\beta_6$ | $\beta_7$ | $\beta_8$ | **66.2** | 28.3 | 10.7 | 5 | 2.3 |
| | i | 0 | $\beta_4$ | $\beta_5$ | $\beta_6$ | $\beta_7$ | $\beta_8$ | 0 | 0 | 0 | 0 | 0 |
| | j | $\beta_3$ | $\beta_4$ | $\beta_5$ | $\beta_6$ | $\beta_7$ | $\beta_8$ | 23.4 | **67.3** | **55.1** | 23.3 | 23.5 |

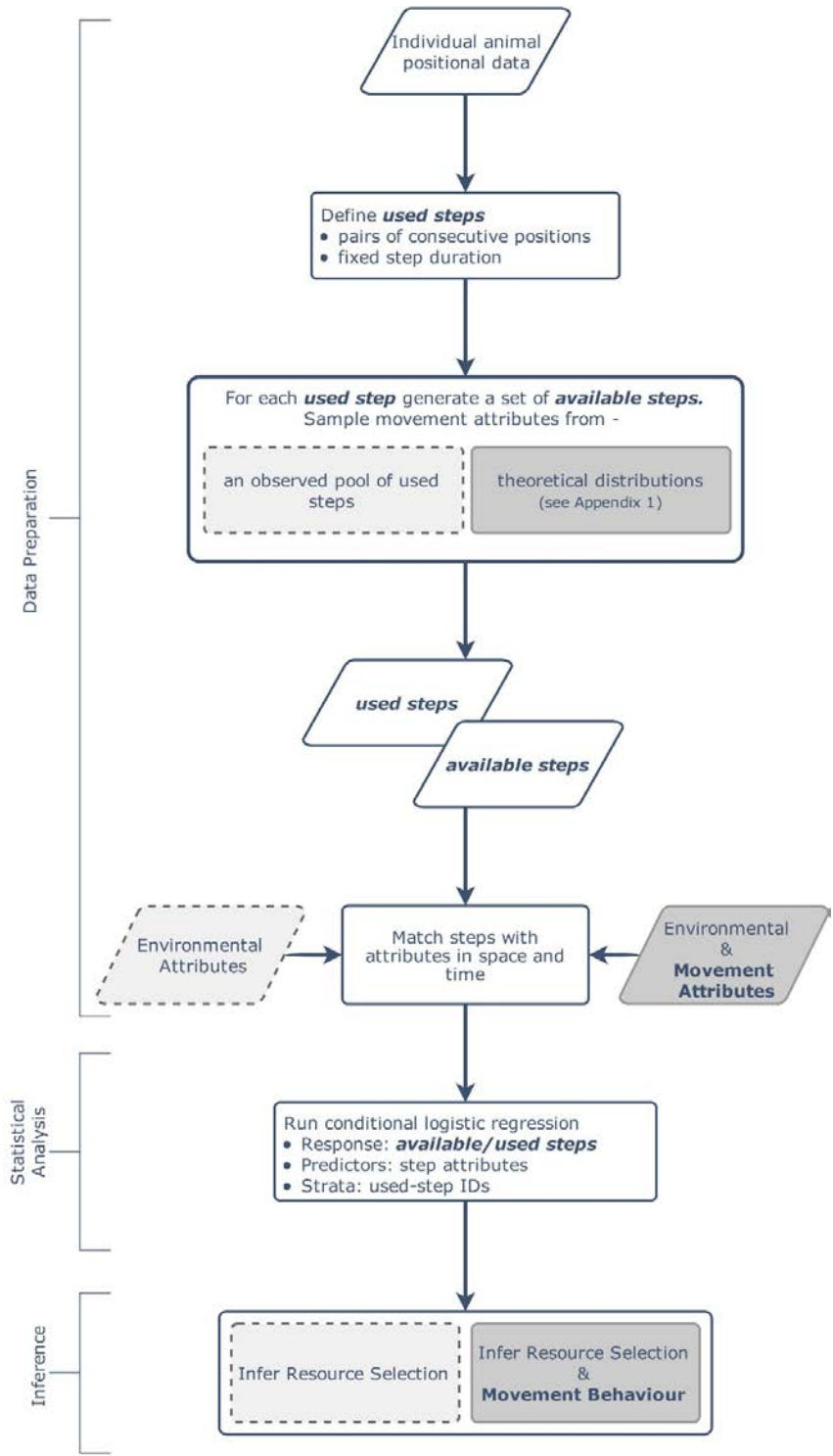

Figure 1 – Step selection analysis workflow. Light grey shading indicates conventional SSA whereas dark grey shading indicates the integrated approach advocated here (iSSA). See Appendix S4 for detailed iSSA guidelines and tips.

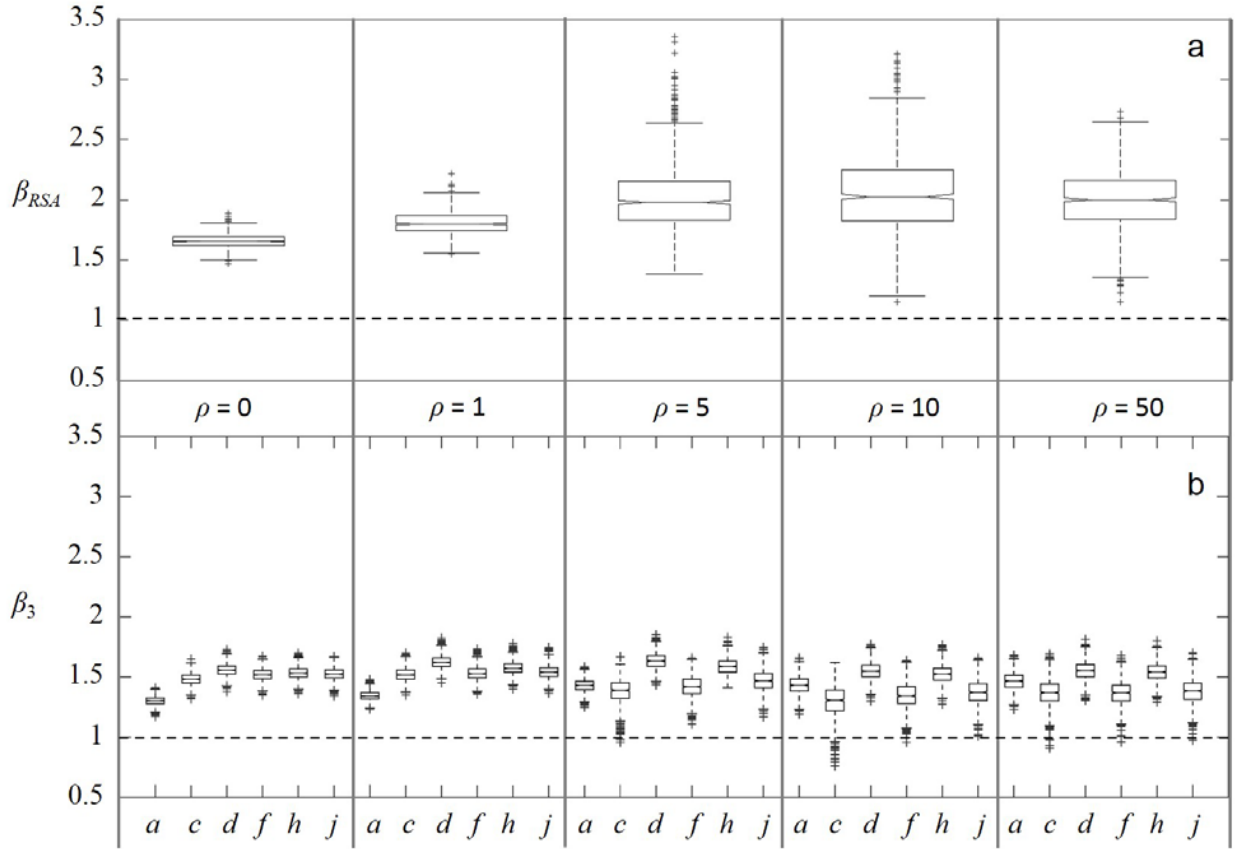

Figure 2 – Statistically inferred habitat selection coefficient estimates for RSA (a) and SSA (b; letters along the x-axis refer to the SSA formulations listed in Table 1), for five levels of habitat spatial autocorrelation, $\rho$. Each box-and-whiskers is based on 1,000 independent estimates, where the central mark is the median, the edges of the box are the 25$^{th}$ and 75$^{th}$ percentiles, the whiskers extend to the most extreme data points not considered outliers (i.e., within approximately 99% coverage if the data are normally distributed), and outliers are plotted individually. Horizontal dashed lines represent the true habitat selection intensity, $\omega = 1$. See Appendix S5 for further details.

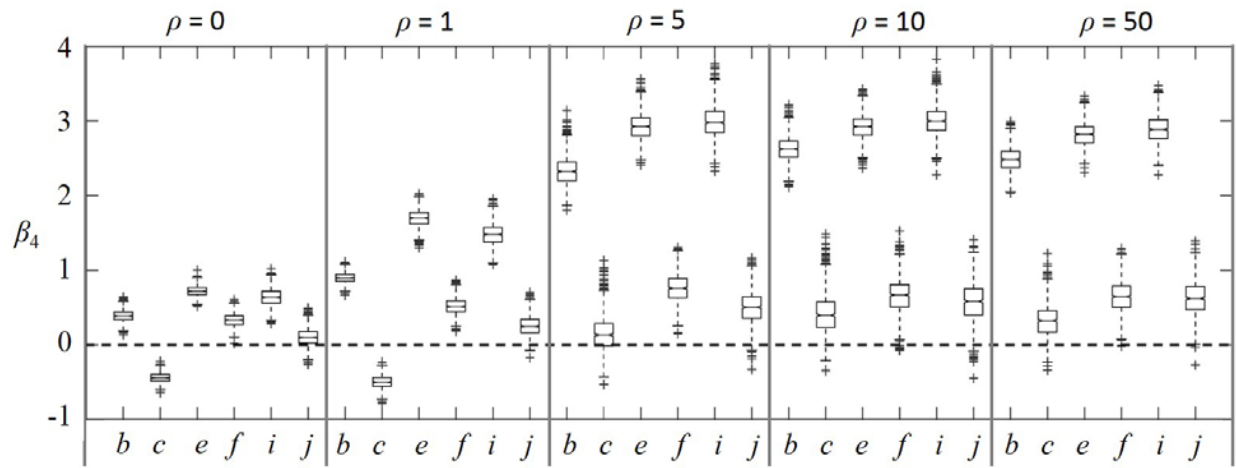

Figure 3 – Statistically inferred effects of the mean habitat along the step. The dashed line represents no effect. Other details are as in Fig 2.

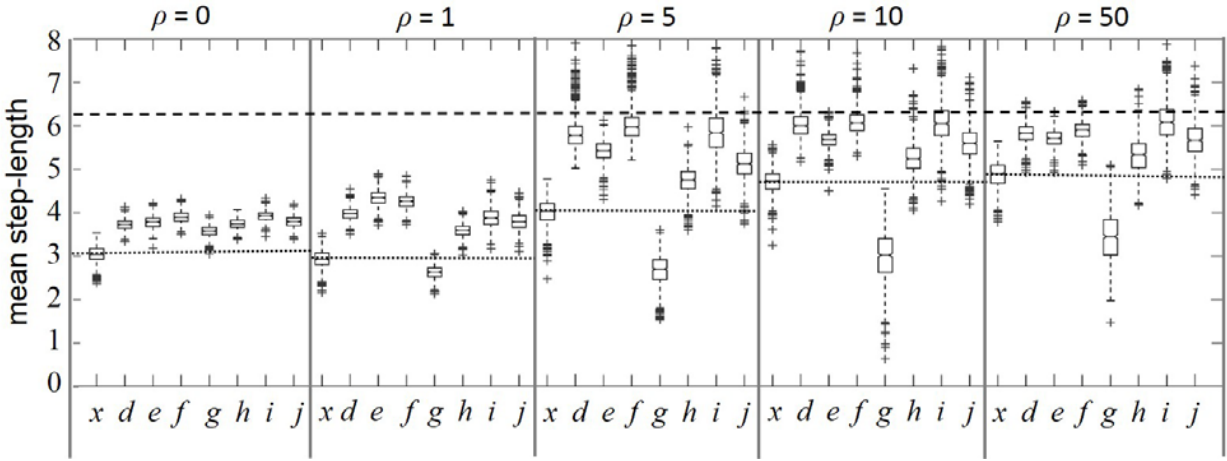

Figure 4 – mean of gamma step-length distributions (displacement in spatial units per $\Delta t$; Appendix S5) inferred based on the different iSSA formulations (see Table 1). Model $x$ is a null model, including only the step-length and its natural logarithm (with no habitat effects), added here to demonstrate that the conditional logistic regression produces unbiased MLEs. The dotted lines correspond to the observed mean step-length across all 1000 realisations at each of the five levels of habitat spatial autocorrelation. The dashed line corresponds to the 'true' habitat-free mean step-length, calculated by simulating the process using Eq S5.1 but with $\omega = 0$. Other details are as in Fig 2.

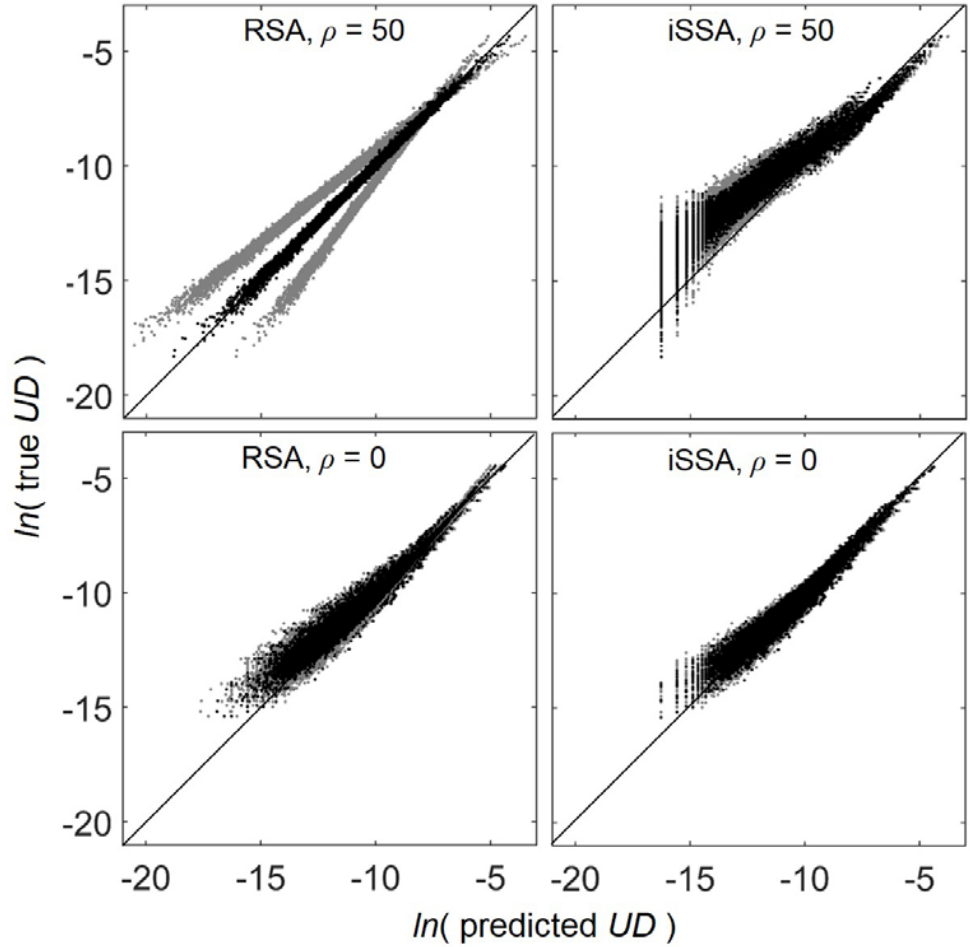

Figure 5 – log-log plots of the true UDs vs the predicted UDs. Each dot represents the utilisation probability of a single spatial cell. Black dots correspond to the median parameter estimates whereas grey dots correspond to the 2.5 and 97.5 percentiles of the estimated parameters distribution. Black diagonal lines represent a perfect 1:1 mapping – dots appearing above these lines are spatial cells where the true UD value exceeded the predicted UD value (under-predictions) whereas dots appearing below these lines represent over-predictions. Note that iSSA results are presented for the simplest iSSA including an endpoint effect, formulation $d$ in Table 1.

**Appendix S1 – from step selection to utilisation distribution**

Step selection models are mechanistic depictions of the movement behavior of individual animals in discrete time. Diffusion approximations allow shifting from this Lagrangian perspective to an Eulerian one. Partial differential equations (Moorcroft & Barnett 2008), or integral equations (Barnett & Moorcroft 2008; Potts *et al.* 2014b), can be used, for example, to approximate some step-selection processes, thus formally linking resource selection and steady-state utilisation distributions (see below). Alternatively, transport equations may provide powerful tools to derive macroscopic patterns from RW-based microscopic movement processes (Kareiva & Odell 1987; Hillen & Painter 2013).

For step-selection processes, the steady-state utilisation distribution is given by the steady state of the following master equation (Okubo & Levin 2001; Potts *et al.* 2014b):

$$UD_{t+\tau}(x') = \int_\Omega R(x'|x)\, UD_t(x)dx, \qquad \text{Eq. 1.1}$$

where $R(x'|x)$ is the redistribution kernel (given by kernel-generating functions such as Eq. 1 or Eq. 4 in the main text) over time interval $\tau$, and $UD_t(x)$ is the utilisation distribution across the spatial domain $\Omega$ at time $t$. The steady-state UD is the limiting function $UD_\infty(x) = \lim_{t \to \infty} UD_t(x)$, which arises as the solution of the following integral equation

$$UD_\infty(x') = \int_\Omega R(x'|x)\, UD_\infty(x)dx. \qquad \text{Eq. 1.2}$$

Eq. 1.2 is usually impossible to solve exactly (but see (Barnett & Moorcroft 2008)). Therefore approximate and/or numerical methods are required. One approximate method involves constructing a partial differential equation (PDE) limit of Eq. 1.1, and performing steady-state analysis (Moorcroft *et al.* 2006; Moorcroft & Barnett 2008). However, such PDEs might not be solvable and/or the approximation might give qualitatively different results to the exact system (Potts & Lewis, in review). Another method involves solving Eq. 1.1 numerically through time until $UD_{t+\tau}(x)$ is sufficiently close to $UD_t(x)$ for all $x$ (Potts *et al.* 2014b). This approach might prove to be particularly computationally demanding due to the computational time required in calculating the integral. To speed this up, one could use Monte-Carlo methods to solve Eq. 1.1, which is equivalent to performing stochastic simulations of the redistribution kernel (as was done in the current study; Appendix S5). After a sufficient burn-in period, the positions of the simulated animals approximately represent samples from the steady-state distribution $UD_\infty(x)$. Regardless of the methods of choice, understanding the relationship between mechanistic kernel-generating functions and resulting UD's is crucial if we wish to translate spatial animal behaviour to population distribution and redistribution patterns.

**Appendix S2 – inferring step-length distributions**

Our aim here is to demonstrate the use of a conditional logistic regression analysis of step-length ($l > 0$) data to obtain maximum likelihood estimates of the parameter/s of one of the following step-length distributions: exponential (with rate $\lambda$), half-normal (with standard deviation $\sigma$), gamma (with shape $k$ and scale $q$), or log-normal (with shape $\sigma$ and scale $\mu$). Taking the natural logarithm of each of these four probability density functions, we obtain:

$$\ln(f_1(l;\lambda)) = \ln(\lambda) - \lambda l, \qquad \text{Eq. 2.1}$$

$$\ln(f_2(l;\sigma)) = -\ln\left(\frac{\sqrt{2}}{\sigma\sqrt{\pi}}\right) - \frac{1}{2\sigma^2}l^2, \qquad \text{Eq. 2.2}$$

$$\ln(f_3(l;k,q)) = -\ln(\Gamma(k)q^k) - \frac{1}{q}l + (k-1)\ln(l), \qquad \text{Eq. 2.3}$$

and

$$\ln(f_4(l;\sigma,\mu)) = \left(-\ln(\sigma\sqrt{2\pi}) - \frac{\mu^2}{2\sigma^2}\right) - \left(\frac{\mu}{2\sigma^2} - 1\right)\ln(l) - \frac{1}{2\sigma^2}\ln(l)^2. \qquad \text{Eq. 2.4}$$

Hence, for all four distributions, the probability of observing a step of length $l$ can be expressed as an exponential function of some linear combination of $l$, $ln(l)$, $l^2$, and/or $ln(l)^2$.

Let us now assume we have obtained a set of $T$ spatial positions, $x_t$, sampled at a unit temporal interval along an animal's path ($t = 1, 2, \ldots, T$), with the $t$'th step-lengths given by $l_t$ ($= \|x_t - x_{t-1}\|$). For simplicity, we shall assume the animal is traversing a homogeneous landscape, that its movement behavior is temporally invariant, and that it lacks any velocity autocorrelation (i.e., the process is first-order Markovian). A step-selection analysis is based on matching each observed point along the path (each but the first), with a set of $s$ control points, $x'_{t,i}$ ($i = 1, 2, \ldots, s$), where the distance between each point and the previous used location is the step-length, $l'_{t,i}$ ($= \|x'_{t,i} - x_{t-1}\|$). We shall start by sampling the set $s$ with equal probability within some maximal distance from $x_{t-1}$ (e.g., a distance corresponding to the maximum movement capacity of the focal species). Using matched case-control logistic regression we now can estimate the value/s of the parameter/s governing the assumed step-length distribution. Assuming an exponential step-length distribution (Eq. 2.1), the likelihood of the observed data is:

$$\prod_{t=2}^{T} \frac{\exp(\beta_1 \cdot l_t)}{\sum_{i=0}^{s} \exp(\beta_1 \cdot l'_{t,i})}, \qquad \text{Eq. 2.1.1}$$

where $\beta_1$ is the statistical estimator of $\lambda$. Similarly, we can formulate the likelihood assuming a half-normal distribution (Eq. 2.2):

$$\prod_{t=2}^{T} \frac{\exp(\beta_1 \cdot l_t^2)}{\sum_{i=0}^{s} \exp(\beta_1 \cdot l'^2_{t,i})}, \qquad \text{Eq. 2.2.1}$$

where $\beta_1$ is the statistical estimator of $\sigma$ ($\beta_1 = -1/2\sigma^2$); the likelihood assuming a gamma distribution (Eq. 2.3):

$$\prod_{t=2}^{T} \frac{\exp(\beta_1 \cdot l_t + \beta_2 \cdot \ln(l_t))}{\sum_{i=0}^{s} \exp(\beta_1 \cdot l'_{t,i} + \beta_2 \cdot \ln(l'_{t,i}))}, \qquad \text{Eq. 2.3.1}$$

where $\beta_1$ is the statistical estimator of $\theta$ ($\beta_1 = -1/q$) and $\beta_2$ is the statistical estimator of $k$ ($\beta_2 = k-1$); and the likelihood assuming a log-normal distribution (Eq. 2.4):

$$\prod_{t=2}^{T} \frac{\exp(\beta_1 \cdot \ln(l_t) + \beta_2 \cdot \ln(l_t)^2)}{\sum_{i=0}^{s} \exp(\beta_1 \cdot \ln(l'_{t,i}) + \beta_2 \cdot \ln(l'_{t,i})^2)}, \qquad \text{Eq. 2.4.1}$$

where $β_2$ is the statistical estimator of $σ$ ($β_2 = -1/2σ^2$) and $β_1$ being the statistical estimator of $μ$ ($β_1 = β_2μ+1$).

The error in these estimators is inversely proportional to *s*. Increased efficiency (i.e., smaller error for the same *s*) can be obtained by sampling the control set under the assumed step-length distribution. To exemplify, assuming an exponential step-length distribution, we first estimate $λ$ based on the observed steps: $λ_1 = \frac{T-1}{\sum_{t=2}^{T} l_t}$. We then sample $T - 1$ sets of *s* control points, $x'_{t,i}$ ($i = 1, 2, …, s$), so that the probability of obtaining a sample at some distance, $l'_{t,i}$, from the previous observed point, is given by the exponential PDF: $λ_1·\exp(-λ_1·l'_{t,i})$. Finally, we get an MLE for $β_1$ using Eq. 2.1.1. Note that, if indeed no other process is considered in the analysis, the expectancy of the resulting $β_1$ is 0. Otherwise, $λ = λ_1 + β_1$. In the main text ('A hypothetical example') and in Appendix S3 we demonstrate this point in the case where sampling is performed under a gamma distribution.

**Appendix S3 – deriving an iSSA likelihood function**

Here we derive Eq. 3 in the main text. We start by deriving the probability density $f(x_t|x_{t-1},x_{t-2})$ of moving to a point $x_t$ at time $t$, given that the previous two positions (at time-steps $t$-1 and $t$-2) were $x_{t-1}$ and $x_{t-2}$ respectively. This probability density is proportional to the product of three expressions, corresponding to propositions A, B and C in the main text (section "A hypothetical example"). Proposition A says that the animal is exponentially selecting for high values of $h(x)$, meaning that $f(x_t|x_{t-1},x_{t-2})$ is proportional to $\exp[\omega \cdot h(x_t)]$ for some $\omega > 0$. Proposition B says that $f(x_t|x_{t-1},x_{t-2})$ is proportional to a von Mises distribution with mean 0 and concentration parameter proportional to $\theta_1+\theta_2 \cdot y(x_{t-1})$. That is, $f(x_t|x_{t-1},x_{t-2})$ is proportional to $\exp([\theta_1+\theta_2 \cdot y(x_{t-1})] \cdot \cos[\alpha_{t-1}-\alpha_t])$, where $\alpha_{t-1}$ and $\alpha_t$ are the headings from $x_{t-2}$ to $x_{t-1}$, and from $x_{t-1}$ to $x_t$, respectively. Finally, proposition C says that $f(x_t|x_{t-1},x_{t-2})$ is proportional to a gamma distribution of the step length, $l_t$ (the Euclidian distance from $x_{t-1}$ to $x_t$). Furthermore, it says that the shape of the distribution depends on the time of day $D_t$, so that $f(x_t|x_{t-1},x_{t-2})$ is proportional to $\exp[\theta_3 \cdot l_t + \ln(l_t) \cdot (\theta_4+\theta_5 \cdot D_t)]$ (see Appendix S2). Taking the product of these three expressions and normalising, we find the following probability density

$$f(x_t|x_{t-1},x_{t-2}) = \frac{\exp[\omega \cdot h(x_t)+[\theta_1+\theta_2 \cdot y(x_{t-1})] \cdot \cos(\alpha_{t-1}-\alpha_t)+\theta_3 \cdot l_t+(\theta_4+\theta_5 \cdot D_t) \cdot \ln(l_t)]}{\int_\Omega \exp[\omega \cdot h(x\prime)+[\theta_1+\theta_2 \cdot y(x_{t-1})] \cdot \cos(\alpha_{t-1}-\alpha_t\prime)+\theta_3 \cdot l_t\prime+(\theta_4+\theta_5 \cdot D_t) \cdot \ln(l_t\prime)]dx\prime}. \quad \text{(Eq 3.1)}$$

Here, $\alpha_t$' is the direction from $x_{t-1}$ to $x$' (any location in the spatial domain), and $l_t$' is the distance between $x_{t-1}$ and $x$'.

The integral in the denominator ensures that $f(x_t|x_{t-1},x_{t-2})$ integrates to 1, so is a probability distribution. This integral can be found numerically, but the calculation is typically very computationally intensive. Therefore, as is standard in SSA, and as explained in the Main Text, we sample $s$ points from an "availability" distribution. In our case, this distribution is given by sampling a direction uniformly at random, and a distance from gamma distribution, $g(l|b_1,b_2)$ (given in the main text in Eq. 2), where $b_1$ and $b_2$ are estimated shape and scale parameters. As explained in Forester *et al.* (2009, Eqs. 4, 5 and surrounding text), this leads to the following discrete-choice approximation for the conditional probability of the animal being at $x$ at time $t$, given its previous two locations and a set of $s$ control points sampled under $g$:

$$\frac{\exp[\omega \cdot h(x_t)+[\theta_1+\theta_2 \cdot y(x_{t-1})] \cdot \cos(\alpha_{t-1}-\alpha_t)+\theta_3 \cdot l_t+(\theta_4+\theta_5 \cdot D_t) \cdot \ln(l_t)]/g(l_t|b_1,b_2)}{\sum_{i=0}^{s} \exp[\omega \cdot h(x'_{t,i})+[\theta_1+\theta_2 \cdot y(x_{t-1})] \cdot \cos(\alpha_{t-1}-\alpha'_{t,i})+\theta_3 \cdot l'_{t,i}+(\theta_4+\theta_5 \cdot D_t) \cdot \ln(l'_{t,i})]/g(l'_{t,i}|b_1,b_2)}, \quad \text{(Eq. 3.2)}$$

where $l'_{t,i}$ are the sampled step lengths, $\alpha'_{t,i}$ are the sampled directions, and $x'_{t,i}$ are the resulting positions found by moving a distance of $l'_{t,i}$ in the direction $\alpha'_{t,i}$.

Substituting $g$ in Eq. 3.2 by its explicit form (Eq. 2 in the main text) yields the following expression for the conditional probability:

$$\frac{\exp[\omega \cdot h(x_t)+[\theta_1+\theta_2 \cdot y(x_{t-1})] \cdot \cos(\alpha_{t-1}-\alpha_t)+(\theta_3+b_2^{-1}) \cdot l_t+(\theta_4-b_1+\theta_5 \cdot D_t) \cdot \ln(l_t)]}{\sum_{i=0}^{s} \exp[\omega \cdot h(x'_{t,i})+[\theta_1+\theta_2 \cdot y(x_{t-1})] \cdot \cos(\alpha_{t-1}-\alpha'_{t,i})+(\theta_3+b_2^{-1}) \cdot l'_{t,i}+(\theta_4-b_1+\theta_5 \cdot D_t) \cdot \ln(l'_{t,i})]} \quad \text{(Eq. 3.3)}$$

from whence Eq. 3 in the main text follows.

**Appendix S4 – iSSA practical guide**

Here we provide tips and guidelines for conducting a fruitful integrated step selection analysis (iSSA). We refer the reader to Thurfjell *et al.* (2014) for a more general review of applications of SSA.

1. <u>Collect animal positional data</u>
    - To maximize the usefulness of the data in an iSSA, positional data should be collected at a constant fix rate (equal time steps).
    - Generally speaking, high fix rate (short time steps and hence short step-lengths) is expected to increase the reliability of the analysis. This is particularly true when aiming to capture continuous use of small spatial units, such as roads. Note however that fix rate should be adjusted to the typical displacement rate of the study species. Fix rate should be considered too high (and hence wasteful) if during a single step the animal is expected to travel less than the positional error (~ 20-30 m for GPS tags), or the spatial resolution of the focal habitat map (~ 20-250 m for most satellite derived maps).
    - Both habitat selection and movement behavior may depend on time of day and season. Sampling should attempt to capture as much temporal variability as possible during the time of the study. For example, fix schedule that is out of synch with time of day (e.g., every 5 hours) can help capture more (temporal variability) with less (fixes).

2. <u>Tabulate observed (case) steps</u>
    - Clean the data - even good GPS datasets contain erroneous positions. Exclude fixes taken before and shortly after tag deployment and after mortality/drop-off events. Plot the observed trajectories and visually look for potential positional errors. Run a code/script that scans the data for extreme values such as unreasonably long (or fast) steps or return-trips (relocations starting and ending at approximately the same point).
    - Because fixes are not always taken at their designated time, it is useful to define a reasonable temporal tolerance range for the step duration (e.g., 1 hr ± 10 min).
    - We recommend including the following fields in the 'used-step' table: individual ID, unique step ID, step start-point (time, easting and northing), step endpoint (time, easting and northing), step-duration, step-length, and step-heading (relative to the true north).
    - If velocity autocorrelation is to be included in the analysis, one must make sure to only calculate velocity deviations (e.g., turn-angles or step-length differences) between successive valid steps. Any step that does not have a valid step leading to it (e.g., because the previous fix is missing), cannot be characterized by valid velocity deviations and hence must be excluded from the analysis (note that such a step should still be used to calculate velocity deviations for the proceeding step).

3. <u>Sample lengths of available (control) steps</u>
    - Available step lengths should be sampled based on one of the following probability density distributions (see Appendix S2): a uniform distribution within some maximal distance (e.g., the longest observed step), exponential, normal, gamma, and log-normal.

- If there is no a priori reason to use a particular distribution, we would recommend using the gamma because it is flexible and includes the exponential as a special case.
- Whatever the theoretical distribution of choice is, the observed step-lengths should be used to estimate its parameters (using either the method of moments or maximum likelihood). If there is strong a priori biological reason to think that these estimates should differ between distinct portions of the data (individual ID, sex, season, study area, etc.), more efficient model fitting can be gained by partitioning the data accordingly.
- If the user is interested in determining which theoretical distribution best fits the data, we recommend using the uniform distribution to sample available step-lengths. The step-length, its natural logarithm, its square, and the square of its natural logarithm should then be included as predictors in a set of four competing models (see Appendix S2 for details), and AIC can be used to choose the best one.

4. Sample available (control) step headings
    - In the simplest case, available step headings are sampled from a uniform distribution between 0 and $2\pi$.
    - If directional correlations or bias are evident in the data, increased efficiency may be gained by sampling available step headings from a von Mises distribution where the concentration parameter is estimated from the observed directional persistence/bias distribution.
    - If observed directional persistence/bias is correlated with step-length (e.g., the animal tends to turn less when making short steps), increased efficiency can be gained by accounting for the correlation structure when sampling available step headings.

5. Generate available (control) steps
    - Combine sampled step lengths and headings (with appropriate cross-correlation structure) to generate available steps starting at a used step start-point and ending in random endpoints.
    - Identify each cluster, consisting of a single used step and its matched set of available steps, with a unique cluster ID. Code all used steps as '1' (case) and all available steps as '0' (control).

6. Attach step attributes
    - Characterize each step (cases and controls) with the following: step-length, step turn-angle and angular deviation from a preferred direction (if relevant), temporal covariates (e.g., time of day, season), and spatial covariates (e.g., elevation, NDVI, cover, temperature, etc.).
    - Covariates could be matched (in space and/or time) to the step's start-point, to its endpoint, and/or based on some interpolation between the two (e.g., average along the step, within an ellipse bounded by the start and end positions, or along a Brownian bridge). Note that, if covariates are purely temporal or are measured at the step start-point, their value would be identical for all steps belonging to the same cluster and their

independent effects are thus statistically unidentifiable. The effects of such covariates are identifiable when interacting with other variables (e.g., an interaction between step-length and season).

7. <u>Fit a conditional logistic regression</u>
   - As long as sample sizes are sufficient (see Appendix S6), we recommend fitting iSSA for each individual independently (rather than using a mixed effects approach). This allows for a straightforward and unbiased evaluation of both inter- and intra-individual variability (Fieberg *et al*. 2010). Population level inference can then be gained by averaging individual model fits.
   - Function *clogit* in R is often used to fit conditional logistic regression. Note that this function (as many other conditional logistic regressions routines) rely on a Cox proportional hazard model to obtain MLEs and hence its output is a *coxph* output.

8. <u>Adjust movement coefficients</u>
   - Once step-length and/or turn-angle coefficient estimates are obtained, those must be combined with the tentative parameter estimates used for sampling available steps (see Appendix 1 and the main text for details). If steps were sampled from a uniform distribution, no adjustments are needed.

9. <u>Simulate space-use</u>
   - Once the integrated step selection function has been fully parametrized, it can be used to simulate space-use across any discrete map of its spatial covariates.
   - This requires a simulation model that iteratively calculates the redistribution kernel at each simulated position and then samples from this kernel to select the next position. Note that the parametrized step-length distribution is one-dimensional, it describes the probability density of any particular displacement over the prescribed time step. When calculating the full two-dimensional redistribution kernel, rather than just drawing from such kernel, the basal probability density for any distance, $r$, from the center of the kernel is proportional to $2\pi r$. One must thus correct for dimensionality by dividing by $2\pi r$. This requires care when the step-length approaches 0, possibly by setting a minimal value for $r$ (e.g., $1/2\pi$), so as to prevent the kernel from collapsing onto a Dirac delta function.
   - A Monte Carlo approximation of the utilization distribution can be gained by simulating long and/or multiple trajectories (starting from the same or different positions) and then normalizing space-use across the map (see also Appendix S5).
   - Such simulations may be used for cross validation (e.g., using ROC AUC to quantify the predictive power of the model over a validation dataset), as well as for predicting the ecological consequences of habitat loss, fragmentation, and other environmental changes that might affect animal space-use.

**Appendix S5 – simulation experiments**

Simulating movement and fitting statistical models: Fine-scale space-use dynamics were stochastically simulated, in discrete time and space, using a 'stepping-stone' movement process. The simulation operates within a discretized spatial domain consisting of 11,600 hexagonal cells (corresponding to 10,000 squared spatial units, so that the distance between adjacent cells is one spatial unit) and wrapped around a torus to eliminate any edge effects. Each cell in the domain, $x$, is characterised by some values of a continuous, normally distributed habitat quality, $h(x)$. Spatial autocorrelation in habitat quality was generated by first assigning a pseudorandom number [~ $U(0,1)$] to each spatial cell in the domain and then locally averaging those within a fixed autocorrelation range, $\rho$. The ranks of the resulting 'smoothed' values were then used to assign $h(x)$ values drawn independently from a standard normal distribution. Five different $\rho$ values were used (0, 1, 5, 10, and 50 spatial units) to generate five levels of spatial autocorrelation (see Fig 5.1 for an illustration, and Appendix S9 for a discussion on the interpretation of $\rho$). For each spatial autocorrelation level, 1,000 different landscape realisations were generated, differing from each other by the spatial configuration of $h$ (but not by its frequency distribution or spatial autocorrelation).

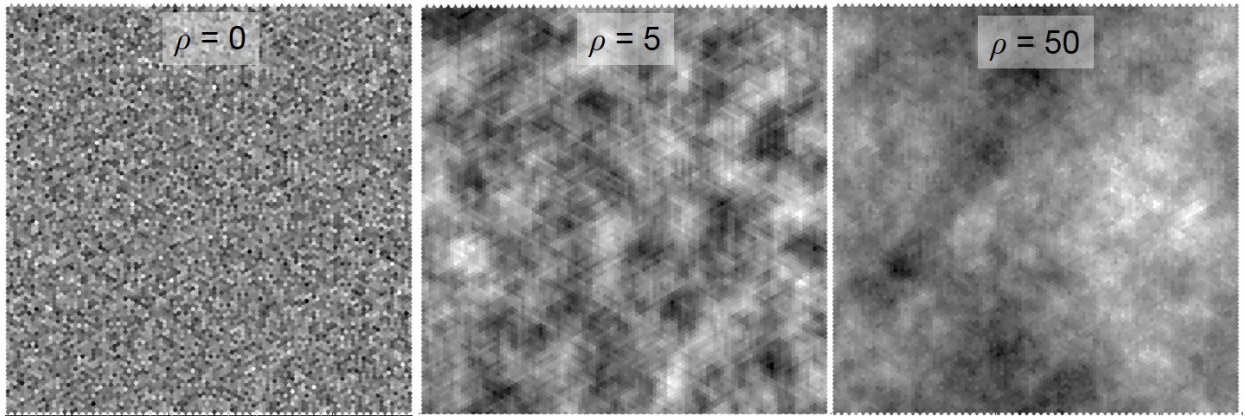

**Figure 5.1 -** Simulated landscape maps with different levels of habitat (grey scale) spatial autocorrelation.

At each simulation time-step, $t$, the position of the animal at the next time-step, $x_{t+\tau}$ (where $\tau$ is the duration of a single simulation step), is stochastically determined according a truncated redistribution kernel, including the current location and the six adjacent cells. The attractiveness of each of the seven available cells is calculated as an exponential function of a basal movement cost, $\mu$, and a habitat induced attraction (or repulsion), a product of local habitat quality, $h(x)$, and the habitat selection intensity, $\omega$. The truncated redistribution kernel is thus given by:

$$p(x_{t+\tau} = x) = \frac{I(\|x-x_t\|\leq 1)\cdot\exp[\omega\cdot h(x)-\mu\cdot\|x-x_t\|]}{\sum I(\|x-x_t\|\leq 1)\cdot\exp[\omega\cdot h(x)-\mu\cdot\|x-x_t\|]}, \qquad \text{Eq. 5.1}$$

where $I$ is an indicator function (valued either 1 or 0, depending on the validity of the expression in parenthesis), and the denominator is a sum over all landscape cells so that the kernel sums to one. Note that here we chose to focus on a stepping-stone process as we aim to model animal movement behaviour at its most fundamental scale, that of a single step. Accordingly, we see the spatial unit of our simulation as approximately equivalent to the body-length of the moving animal, and the temporal unit, $\tau$, as approximately equivalent to the time required to move one such spatial unit. Also note that, for simplicity, we did not include any directionality effects (neither persistence nor bias) in the simulation nor in the subsequent analysis (we refer the reader to Duchesne *et al.* 2015 for further details on this topic).

Equation 5.1 was used to generate 1,000 'observed' positional time series for each of the five spatial autocorrelation levels – one trajectory was simulated across each of the 1,000 landscape realisations for each of the five $\rho$ values. Unless otherwise specified, all simulations were run with $\omega = 1$ and $\mu = 1.7918$ (corresponding to a habitat-independent movement probability of 0.5). Simulated trajectories were always initiated in the best cell in each landscape (the one with the highest $h$ value). The first 20,000 (approximately twice the domain size) time steps of each trajectory were omitted as a 'burn-in period', whereas the remaining $10^5$ time steps were rarefied by retaining every 100$^{th}$ position, resulting in an 'observed' time-series of $T = 1,000$ positions (with observed step-duration, $\Delta t = 1 = 100 \cdot \tau$). Rarefication was performed here so as to emulate typical telemetry data where most animals likely adjust their spatial position at a substantially higher rate (say, every few minutes or seconds) than the rate at which we sample their position (typically, every few hours).

Each of the resulting 5,000 rarefied trajectories were then separately analysed using RSA and SSA formulations of various levels of complexity (including several iSSAs), and maximum-likelihood coefficient estimates were obtained for each realisation. For the RSA, occurrences were coded as 1 ('used') whereas all cells in the landscape were coded as 0 ('available'; see Discussion for implications). The resulting binary response variable was then statistically modeled as function of the local habitat value, $h(x)$, using logistic regression (function *glm* in R with a Binomial eror distribution and a logit link). This yielded, for each observed trajectory, two RSA-based coefficients estimates, an intercept (a meaningless scaling parameter in the case of our 'used vs available' design), and a selection coefficient, $\beta_{RSA}$.

For the SSAs, each observed point along each rarefied trajectory (but the first; $x_t$, $t = 2, 3, \ldots, T$) was matched with an availability set of $s = 10$ random spatial positions ($x'_{t,i}$, $i = 1, 2, \ldots, 10$). The probability of sampling an available point was a sole function of its distance from the previously observed point (i.e., the length of the potential step ending at that point, $l'_{t,i}$), and was given by a gamma distribution with $\rho$-specific shape and scale parameters ($\beta_1$ and $\beta_2$) estimated from the joint distribution of observed step-lengths across all 1,000 realisations. To enable log-transformations and for consistency with kernel generating (see Appendix S4, section 9), the length of all steps (used and available) that resulted in 0 displacement (staying in the same cell) was set to $1/2\pi$. As in the RSA, used (i.e., observed) points were coded as 1 ('case') whereas available points were coded as 0 ('control'). The resulting binomial response variable was then statistically linked to various covariates using conditional (case-control) logistic regression (function *clogit* in R, with point ID as the strata), fitting an independent model to each trajectory.

Ten different SSA formulations were fitted using one or more of the following six covariates (see Table 1): habitat values at the end of each step, $h(x_t)$, the average habitat value along each step, $h(x_{t-1},x_t)$, the step-length, $l_t$ ($= ||x_{t-1} - x_t||$), its natural-log transformation, $ln(l_t)$, and the statistical interactions between $l_t$, $ln(l_t)$, and $h(x_{t-1},x_t)$. Note that 'steps' are treated here as straight line segments along which the habitat is averaged. This is the most common formulation found in the literature but is certainly not the only one. Steps may be instead defined, for example, as ellipses bounded by two consecutive positions, or even as a Brownian bridge (a spatial probability density kernel derived from an explicit diffusion process). Models that included only $h(x_t)$ and/or $h(x_{t-1},x_t)$ correspond to traditionally used SSA (models *a*, *b*, and *c* in Table 1), whereas models that additionally included $l_t$ and $ln(l_t)$ correspond to iSSA. The likelihood of a single observed trajectory given the full model (including all six covariates; model *j* in Table 1) is exactly proportional to:

$$\prod_{t=2}^{1000} \frac{\exp[\beta_3 \cdot h(x_t) + \beta_5 \cdot h(x_{t-1},x_t)] + [\beta_5 + \beta_7 \cdot h(x_{t-1},x_t)] \cdot l_t + [\beta_6 + \beta_8 \cdot h(x_{t-1},x_t)] \cdot \ln(l_t)]}{\sum_{i=0}^{10} \exp[\beta_3 \cdot h(x'_{t,i}) + \beta_4 \cdot h(x_{t-1},x'_{t,i}) + [\beta_5 + \beta_7 \cdot h(x_{t-1},x'_{t,i})] \cdot l'_{t,i} + [\beta_6 + \beta_8 \cdot h(x_{t-1},x'_{t,i})] \cdot \ln(l'_{t,i})]},$$ Eq. 5.2

where the 0'th available step correspond to the used step ($x'_{t,i=0} = x_t$; note the similarity to Eq. 3 in the main text). The derivation of Eq. 5.2 is identical in form to that of Eq. 3 (see Appendix S3) so is omitted here. This formulation allows for statistically inferring a hypothetical movement process with gamma-distributed step-lengths (with shape and scale that could be governed by the traversed habitats), a habitat-mediated step selection, and a habitat-mediated destination (i.e., endpoint) selection.

Quantifying and comparing utilisation distributions: The predictive capacity of the models was estimated based on the agreement between their predicted utilisation distributions (UD) and the 'true' UD, generated by the true underlying movement process (i.e., Eq. 5.1). UDs were generated across a single landscape realisation for each of the five $\rho$ values ('validation landscapes', independent of those used to generate the movement data), where the same validation landscape was used to generate true-UDs, RSA-based UDs, and SSA-based UDs. The RSA-based UD value at each landscape cell, *x*, was calculated as $\frac{exp(\widetilde{\beta_{RSA}} \cdot h(x))}{\sum_{x=1}^{11600} exp(\widetilde{\beta_{RSA}} \cdot h(x))}$, with $\widetilde{\beta_{RSA}}$ being the median of the 1,000 $\beta_{RSA}$ estimates obtained for any particular $\rho$ value (the median was used here as a measure of centrality that is relatively insensitive to outliers).

For the movement models (the true process, given by Eq. 5.1, and the step-selection process parameterized based on the various SSAs), steady-state utilisation distributions were generated by simulating 11,600 trajectories, each starting from a different cell across each of the five validation landscapes (see Appendix S1 for alternative approaches). For the true UDs, these simulations were carried out using Eq. 5.1, with the same parameter values used to generate the original simulated trajectories ($\omega = 1$ and $\mu = 1.7918$). As before, the first 20,000 time-steps of each trajectory were omitted as a 'burn-in period' and the remaining $10^5$ time-steps were retained. The number of time steps spent in each cell in the domain (across all 11,600 simulated trajectories) was then divided by $1.16 \cdot 10^9 \, \tau$ (the total time spent in the domain by all simulated individuals) to yield the true UD for a single landscape realisation of a given $\rho$ value. SSA-based UDs were generated in a similar fashion to the true UDs, but using the parameterized step-selection models instead of Eq. 5.1, with parameter values corresponding to the median of the 1,000 $\beta_1$ estimates obtained for any particular $\rho$ value. To match the temporal extent of true UD simulations ($\Delta t = 100 \cdot \tau$), SSA simulations were run for 1,200 steps and the initial 200 steps where omitted as a burn-in period.

Thus far we have described the generation of (approximately) steady-state UDs, pertaining to the population's spatial distribution as measured over extended time periods, and assumed to be temporally stable. However, for our movement models it also is possible to calculate and compare transient UDs by simulating movement over shorter time spans. The shorter the time span, the more sensitive the UD is to initial conditions (i.e., the path's starting point). Transient UDs were simulated similarly to the 'steady-state' UDs, except for three differences. First, simulated trajectories were always initiated in the best cell in the landscape (the one with the highest *h* value). Second, no burn-in period was omitted for the SSA-based simulations, and only the first 99 steps were omitted from the true simulations (to match the SSA starting conditions). Third, each path lasted only $10^4 \cdot \tau$ (or $100 \cdot \Delta t$ in the SSA case). Other details are as above.

To compare the true-UDs (*truth*) to the statistically predicted ones (*model*) we use the Kullback-Leibler Divergence, *KLD*(*model*,*truth*), a measure of the information lost when the latter is used to approximate the former (see Potts *et al.* 2014c for similar usage). Note that the 'earth mover's distance' (Rubner *et al.* 2000), also may be a useful tool if one wishes to analyse the detailed movement dynamics rather than the resulting UD, as it can give various aspects of information regarding model bias and predictive power (Potts *et al.* 2014a). To facilitate intuitive interpretation of the resulting values, we further define a *KLD*-based performance measure (i.e., a pseudo $R^2$), $G_{KLD}$, quantifying goodness-of-fit relative to an information-free null model – a uniform UD (*null*):

$$G_{KLD}(model, null, truth) = e^{-KLD(model,truth)/KLD(null,truth)} \qquad \text{Eq. 5.3}$$

Hence, the better the focal model performs compared to a null expectation, the closer its $G_{KLD}$ value would be to 1, whereas if it performs worse than the null, its $G_{KLD}$ value would drop below ~0.368 and asymptotically approach 0.

**Appendix S6 – Evaluating the iSSA parameter identifiability and estimability**

To evaluate the identifiability and estimability of the iSSA parameters we simulated three, $10^6$ steps-long, trajectories across the intermediate spatial autocorrelation ($\rho = 5$) validation landscape (see Appendix S5 for details about landscape configurations). Trajectories were generated by stochastic sampling out of a redistribution kernel of the general form of Eq. 1 in the main text.

Methods: The movement kernel was defined by a uniform-random turn distribution (i.e., no directional persistence or bias) and gamma distributed step-lengths. The shape, $k$, and scale, $q$, of the step length distribution varied across the landscape as functions of the habitat value at the step's start point, $h(x)$:

$$k(x) = 2 \cdot \left[1 + \frac{h_{max} - h(x)}{h_{max} - h_{min}}\right], \qquad \text{Eq. 6.1}$$

and

$$q(x) = 5 \cdot \left[1 + \frac{h(x) - h_{min}}{h_{max} - h_{min}}\right]^{-1}, \qquad \text{Eq. 6.2}$$

where $h_{max}$ and $h_{min}$ are the maximal and minimal habitat values occurring on the landscape. Hence, $k$ is a linear function of $h$ with intercept $= 2\left(1 + \frac{h_{max}}{h_{max}-h_{min}}\right)$ and slope $= \frac{-2}{h_{max}-h_{min}}$; as $h$ increases, $k$ decreases from 4 to 2. Similarly, $q^{-1}$ is a linear function of $h$ with intercept $= 0.2\left(1 - \frac{h_{min}}{h_{max}-h_{min}}\right)$ and slope $= \frac{0.2}{h_{max}-h_{min}}$; as $h$ increases, $q$ decreases from 5 to 2.5. The habitat selection function was given by $exp[\omega \cdot h(x)]$ with one of three values of the selection coefficient, $\omega = 0$, 1, or 2 (hence the three different trajectories). Biologically, this scenario may correspond, for example, to habitat-induced foraging behavior where animals tend to move less as their local habitat value increases (area restricted search), while simultaneously selecting locations with high habitat value.

As described in Appendix S5, each used point (but the first; $x_t$, $t = 2, 3, …, 10^6$) along each of the three trajectories was matched with an availability set of $s = 10$ random spatial positions ($x'_{t,i}$, $i = 1, 2, …, 10$). The probability of sampling an available point was a sole function of its distance from the previously observed point (i.e., the length of the potential step ending at that point, $l'_{t,i}$), and was given by a gamma distribution with shape and scale parameters ($\beta_1$ and $\beta_2$) estimated based on all observed step-lengths along each trajectory. Used points were coded as 1 ('case') whereas available points were coded as 0 ('control'). The resulting binomial response variable was then statistically linked to the following covariates using conditional (case-control) logistic regression (function *clogit* in R with point ID as the strata): the habitat values at the end of each step, $h(x_t)$, the step-length, $l_t$, its natural-log transformation, $ln(l_t)$, and the products of the habitat value at the step's start point, $h(x_{t-1})$ and the step length and its natural-log transformation, $l_t \cdot h(x_{t-1})$ and $ln(l_t) \cdot h(x_{t-1})$.

To evaluate identifiability, parameters were estimated based on the entire trajectory (999,999 steps; this takes 5-10 CPU minutes). To evaluate estimability under constrained sample size, the model was fitted 1,000 times, each time with a different segment of the full trajectory, where segment length varied from 100 to 1,000 observed points (and hence sample size varied from 99 to 999 steps).

Results and discussion: All model fits converged successfully and in a timely manner. Parameter estimates obtained based on the full trajectory were unbiased (i.e., accurate) compared to the true values. Moreover, parameter estimates obtained based on shorter segments of the trajectory were also, on average, unbiased compared to the true values.

A model may be considered unidentifiable if the maximum of the likelihood surface occurs at more than just a single point in the parameter space, often along a 'ridge' (a line or a line segment through the parameter space along which the likelihood is exactly the maximum likelihood). Where the likelihood function is analytically tractable, detecting model unidentifiability is simply a question of whether the maximum of the likelihood function is unique. Most often however, numerical approximations are used to obtain MLE's, approximations that may fail to detect a ridge in the likelihood surface, and hence an unidentifiable model. We thus rely here on two indicators of model identifiability. First, we rerun the model fit 10 times for each of the three trajectories, making sure that parameter estimates did not change between fits (to the $8^{th}$ significant figure). Second, we used the standard errors of the parameter estimates to assess the steepness of the likelihood profile around the MLEs, where large standard errors (despite the very large sample size) would have been indicative of a flat likelihood profile and hence a potential ridge. All standard errors were very small (relative to the magnitude of the parameter estimates) and we thus conclude all five parameters in this scenario are identifiable, regardless of the strength of habitat selection (Figs 6.1-6.3).

Identifiability is a necessary but not sufficient condition for estimability. To evaluate estimability we focus here on the precision of parameter estimates obtained from multiple independent realizations of the same process. Low precision means that different realizations of the same process yield markedly different parameter estimates, indicating an estimability problem. As can be expected, precision (and hence estimability) increased with sample size, but with diminishing returns beyond ~400 observed positions (Figs 6.1 – 6.3). For the smallest sample size used here (99 steps) precision was very poor, particularly for the parameters related to the scale of the step-length distribution. Estimates of the intercept of the shape function, and both the intercept and the slope (with regards to *h*) of the inverse scale function, show decreased precision as selections strength increases, a pattern not observed for the other two parameters.

Low precision, particularly when sample size is low, was likely driven by high correlations between some of the statistical coefficients (Table 6.1). The movement components of the iSSA are inherently 'correlation-prone' and are hence vulnerable to estimability issues. Whereas the current analysis indicates satisfactory inferential performance, the results may differ under different scenarios. It is thus important to note that estimability analysis should be tailored to the specific system and model that are being evaluated. This is particularly true when additional movement components are included in the model, such as directional persistence and/or bias, where strong collinearity may result in loss of estimability. We highly recommend users to take advantage of the mechanistic nature of the iSSA and simulate potential space-use patterns under the desired model structure so that estimability can be adequately assessed.

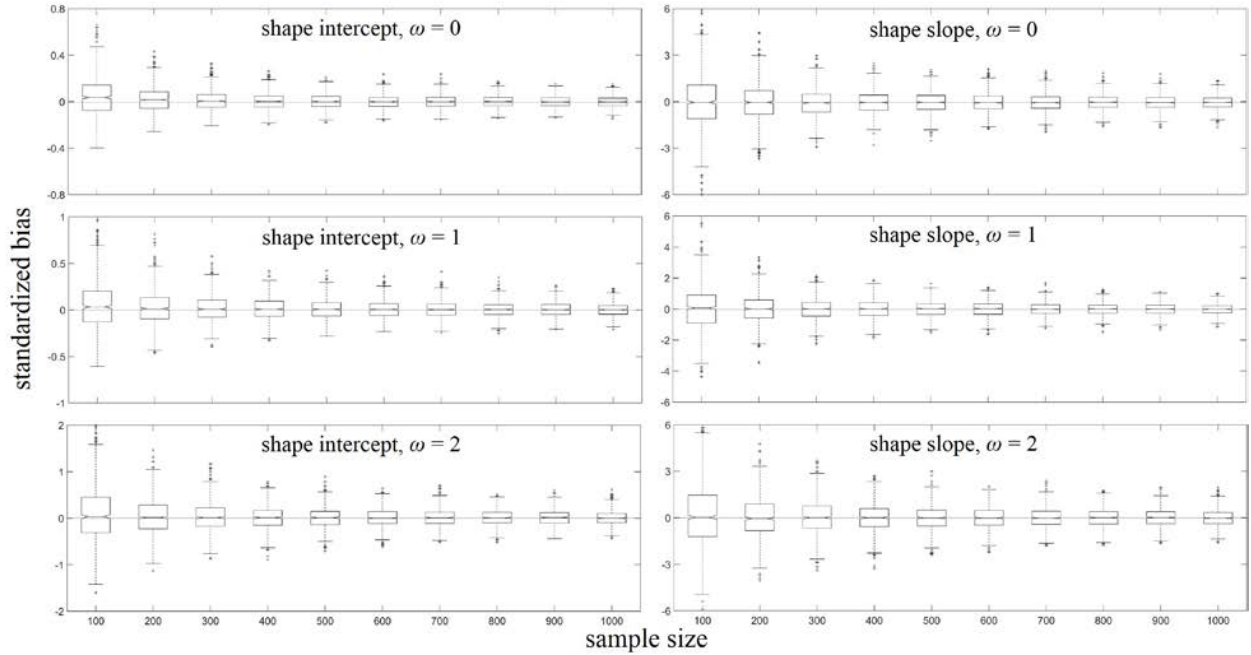

**Figure 6.1** – the standardized bias (the difference from the true value divided by that true value) in the estimated linear intercept and slope (the effect of *h*) of the shape parameter of the step-length distribution as function of sample size. Grey lines reflect the 95% confidence region of the estimates obtained based on the full trajectory (999999 steps; based on the estimated standard errors) and is hence a measure of identifiability.

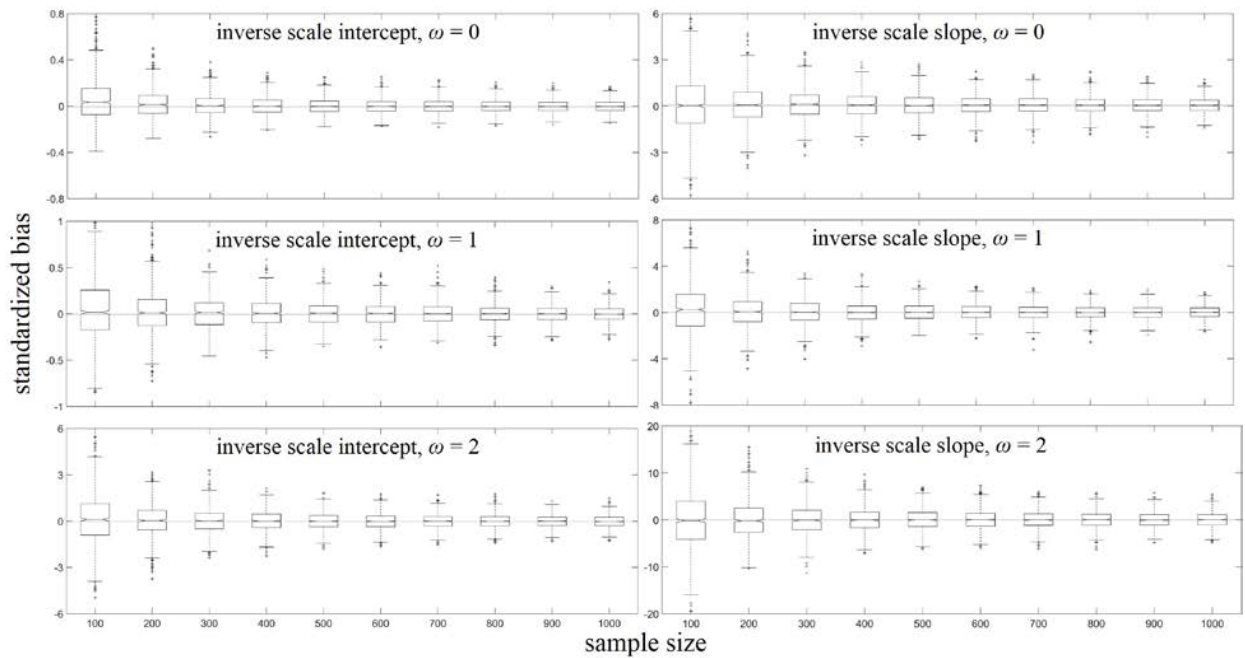

**Figure 6.2** – the standardized bias in the estimated linear intercept and slope of the inverse scale parameter of the step-length distribution as function of sample size. Grey lines reflect the 95% confidence region of the estimates obtained based on the full trajectory.

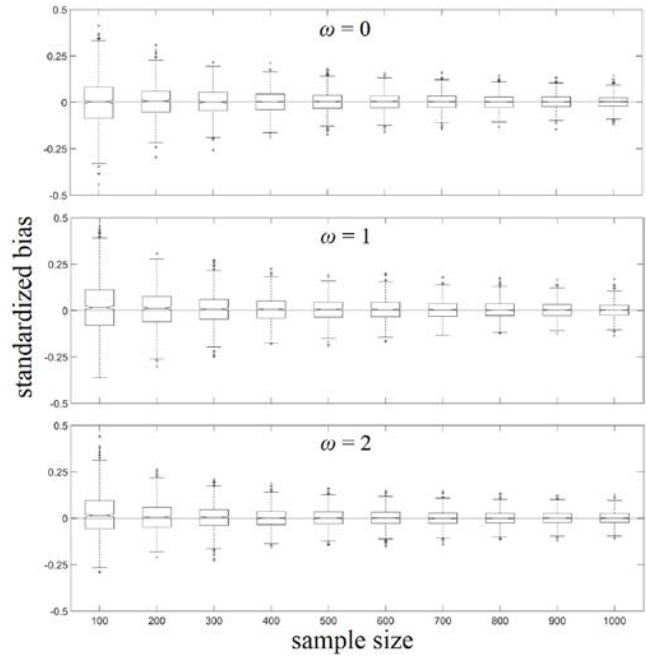

**Figure 6.3** – the standardized bias in the estimated selection coefficient as function of sample size. Grey lines reflect the 95% confidence region of the estimates obtained based on the full trajectory.

| covariate 1 | covariate 2 | Pearson's $r$ | | |
|---|---|---|---|---|
| | | $\omega = 0$ | $\omega = 1$ | $\omega = 2$ |
| $h(x_t)$ | $l_t$ | 0.067 | 0.001 | 0.111 |
| | $ln(l_t)$ | -0.074 | 0.016 | -0.128 |
| | $l_t \cdot h(x_{t-1})$ | -0.120 | 0.021 | -0.049 |
| | $ln(l_t) \cdot h(x_{t-1})$ | 0.191 | 0.063 | 0.164 |
| $l_t$ | $ln(l_t)$ | -0.914 | -0.874 | -0.833 |
| | $l_t \cdot h(x_{t-1})$ | 0.071 | -0.715 | -0.959 |
| | $ln(l_t) \cdot h(x_{t-1})$ | 0.133 | 0.782 | 0.840 |
| $ln(l_t)$ | $l_t \cdot h(x_{t-1})$ | 0.054 | 0.543 | 0.739 |
| | $ln(l_t) \cdot h(x_{t-1})$ | -0.270 | -0.865 | -0.981 |
| $l_t \cdot h(x_{t-1})$ | $ln(l_t) \cdot h(x_{t-1})$ | -0.867 | -0.778 | -0.789 |

**Table 6.1** – pairwise Pearson's correlations among model coefficients. Correlations were calculated based on the 1000 repetitions of the model fit with a sample size of 999.

**Appendix S7 - $\beta_7$ and $\beta_8$**

Statistically inferred interactions between the mean habitat along the step and the step-length (a) and the natural logarithm of the step-length (b). The dashed line represents no effect. Other details are as in Fig 2 in the main text.

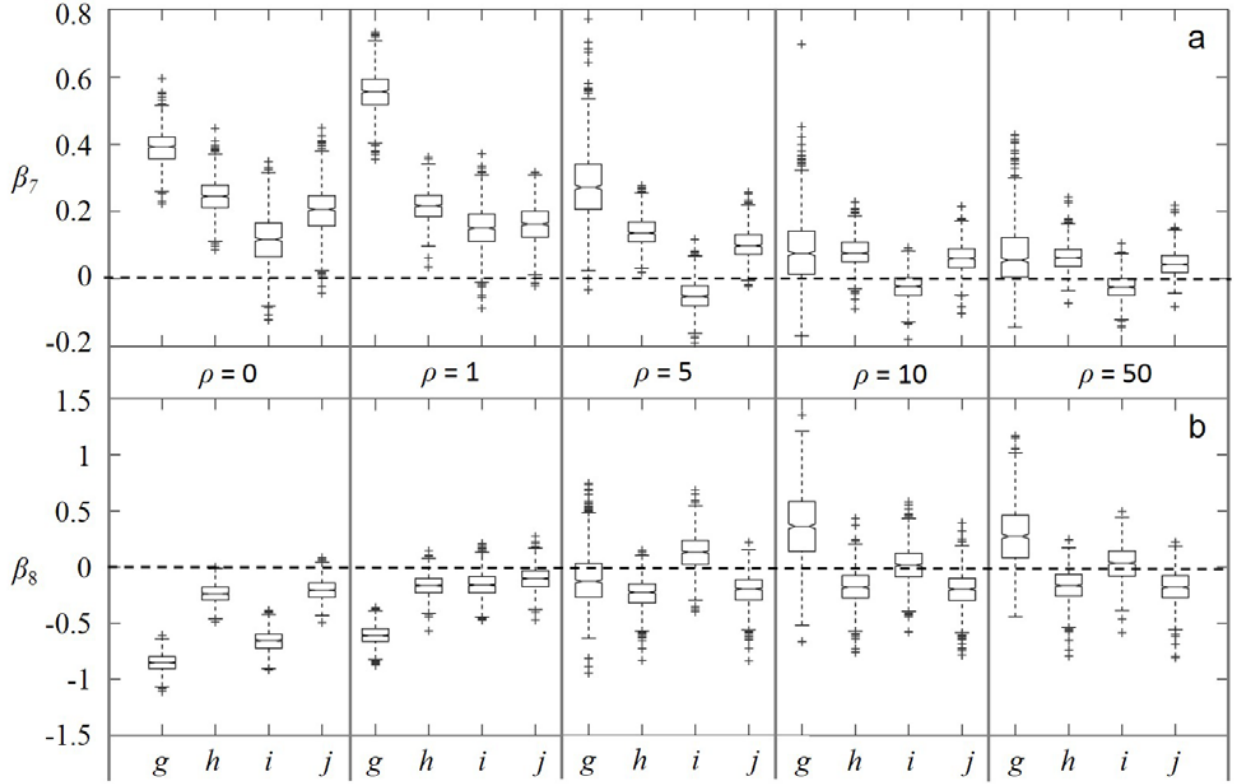

**Appendix S8 – predictive performance**

The 11 different models evaluated here (10 SSAs + 1 RSA) and their Kullback-Leibler divergence from the true UDs (steady-state and transient) at five different levels of habitat spatial autocorrelation. SSA formulations including an endpoint effect ($\beta_3$) are highlighted in grey. Models $d$ to $j$ include step-length effects (iSSA). The lowest (best-performing) SSA Kullback-Leibler divergence values at each level of habitat spatial autocorrelation are bolded to allow comparison with the corresponding RSA values. For reference, the first row lists the KLD values for null model - a uniform UD. Note the superior performance of the RSA projections in predicting the steady-state UDs. Interestingly, a similar result was obtained by Fieberg (2007), who found that ignoring the autocorrelation in the positional time-series (as is done in RSA) often led to slightly better estimates of the UD.

| model | KLD (steady state) | | | | | KLD (transient) | | | | |
|---|---|---|---|---|---|---|---|---|---|---|
| | $\rho=0$ | $\rho=1$ | $\rho=5$ | $\rho=10$ | $\rho=50$ | $\rho=0$ | $\rho=1$ | $\rho=5$ | $\rho=10$ | $\rho=50$ |
| null | 1.48 | 1.66 | 1.80 | 1.78 | 1.84 | 2.78 | 3.62 | 3.00 | 2.66 | 2.27 |
| RSA | 0.02 | 0.02 | 0.01 | 0.02 | 0.01 | 1.02 | 1.28 | 0.65 | 0.34 | 0.07 |
| $a$ | 0.28 | 0.19 | **0.08** | **0.08** | **0.12** | 2.52 | 1.05 | 0.10 | **0.06** | 0.09 |
| $b$ | 1.14 | 0.80 | 0.26 | 0.24 | 0.18 | 1.61 | 0.76 | 0.28 | 0.26 | 0.21 |
| $c$ | 0.32 | 0.19 | 0.10 | 0.11 | 0.19 | 2.09 | 0.73 | 0.14 | 0.08 | 0.13 |
| $d$ | **0.02** | **0.09** | 0.14 | 0.13 | 0.16 | **0.08** | **0.07** | **0.09** | 0.08 | **0.09** |
| $e$ | 1.05 | 0.92 | 0.28 | 0.29 | 0.21 | 0.95 | 1.17 | 0.29 | 0.29 | 0.21 |
| $f$ | 0.03 | 0.15 | 0.31 | 0.19 | 0.33 | 0.12 | 0.13 | 0.14 | 0.11 | 0.18 |
| $g$ | 0.52 | 0.36 | 0.95 | 1.42 | 1.55 | 4.42 | 3.03 | 0.82 | 1.25 | 0.92 |
| $h$ | 0.19 | 0.12 | 0.18 | 0.11 | 0.22 | 0.32 | 0.09 | 0.10 | 0.08 | 0.12 |
| $i$ | 0.78 | 0.79 | 0.30 | 0.27 | 0.21 | 4.49 | 0.93 | 0.28 | 0.27 | 0.21 |
| $j$ | 0.05 | 0.15 | 0.22 | 0.18 | 0.19 | 0.13 | 0.13 | 0.11 | 0.12 | 0.11 |

**Appendix S9 – interpreting $\rho$**

Statistical inference of animal movement behaviour is highly scale-dependent, varying with both the temporal resolution of the observed positional time series and the spatial resolution (or grain) of the landscape. For example, Avgar *et al.* (2013) investigated the sensitivity of parameter estimates in their cognitive-movement model to coarsening (or rarefying) both temporal and spatial resolutions of their simulated data. Here we evaluate these scale-sensitivities by varying a single parameter, $\rho$, the landscape's spatial autocorrelation range. Note however that $\rho$ be interpreted in various ways. The most straightforward interpretation is to think of $\rho$ as a measure of habitat patchiness, ranging from no patches ($\rho = 0$), through multiple small patches, to a single patch covering (on average) half of the spatial domain ($\rho = 50$). Alternatively, $\rho$ may be thought of as a being inversely related to the gap between the scale of the behavioural process and the scale of the observation. In between observed positions (in our case, 99% of the time), a moving animal experiences substantially more environmental variability as $\rho$ declines from 50 to 0, variability that is not observed in the rarefied data. Hence, increasing $\rho$ can be interpreted as decreasing rarefication, and its effect on parameter estimates and predictive capacity can be viewed in this light. Lastly, $\rho$ can be interpreted as a measure of the animal's movement (or patch departure) rate. During the observation period, an animal may visit many different habitat patches ($\rho = 0$), or stay in a single one ($\rho = 50$), depending on a variety of factors (e.g., the ratio between the animal's metabolic requirements and the patch's productivity). Instead of explicitly modeling these factors, $\rho$ can be thought of as crudely representing the resulting spatiotemporal relationships between the animal and the grain of the landscape. For an in-depth discussion of spatial scale dependency we refer the reader to Sandel (2015).

**Appendix S10 – habitat selection and utilisation distribution**

The natural logarithm of utilisation distributions emerging from a simple stepping-stone movement process with exponential habitat selection at four levels of spatial autocorrelation, plotted against the underlying habitat value at each cell (the fifth level, $\rho = 50$, was omitted as it is indistinguishable from $\rho = 10$). Grey lines represent best fit linear regression, with their parameter estimates presented on the top left corner of each panel. Note that our spatially discrete simulation model does not allow the scale of the movement to exceed the scale of habitat variation, only to equal it ($\rho = 0$), and as a result, the observed log-linear slope does not approach 1 but is rather minimised at an intermediate value.

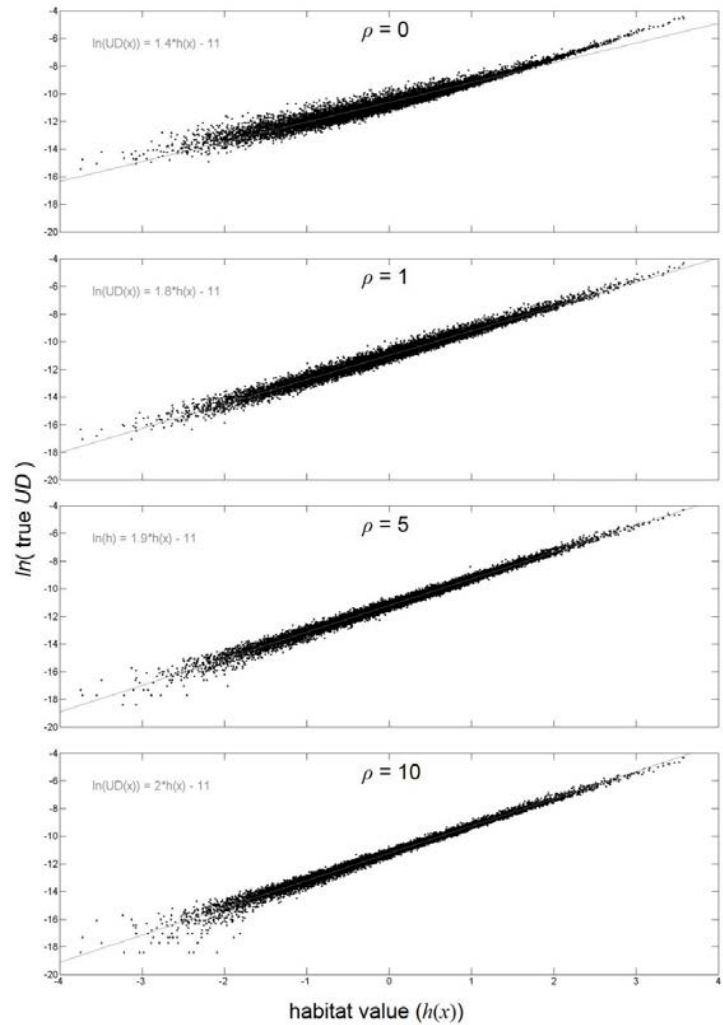

**Appendix S11 – appendices reference list**